\documentclass[manuscript,screen,review=false]{acmart}

\AtBeginDocument{%
  }

\setcopyright{acmcopyright}
\copyrightyear{2018}
\acmYear{2018}
\acmDOI{XXXXXXX.XXXXXXX}

\acmISBN{978-1-4503-XXXX-X/18/06}

\usepackage{subcaption}
\usepackage{multirow}
\usepackage{listings}

\begin{document}

\title{Relevant information in TDD experiment reporting}


\author{Fernando Uyaguari}
\affiliation{%
  \institution{Instituto Superior Tecnológico Wissen}
  \streetaddress{Jose Ma. Sanchez y Av. 10 de Agosto}
  \city{Cuenca}
  \country{Ecuador}
  \postcode{010107}}
\email{fernando.uyaguari@wissen.edu.ec}

\author{Silvia T. Acu\~{n}a}
\affiliation{%
  \institution{Escuela Politécnica Superior, Universidad Autónoma de Madrid}
  \streetaddress{Calle Francisco Tomás y Valiente 11}
  \city{Madrid}
  \country{Spain}
  \postcode{28049}
}
\email{silvia.acunna@uam.es}

\author{John W. Castro}
\affiliation{%
 \institution{Departamento de Ingeniería Informática y Ciencias de la Computación, Universidad de Atacama}
 \city{Copiapó}
 \country{Chile}}
\email{john.castro@uda.cl}

\author{Davide Fucci}
\affiliation{%
  \institution{Blekinge Institute of Technology}
  \country{Sweden}}
\email{davide.fucci@bth.se}

\author{Oscar Dieste}
\affiliation{%
  \institution{Escuela Técnica Superior de Ingenieros Informáticos, Universidad Politécnica de Madrid}
  \streetaddress{Campus Montegancedo, Calle ciruelos, s/n}
  \city{Boadilla del Monte, Madrid}
  \country{Spain}
  \postcode{28660}}
\email{odieste@fi.upm.es}

\author{Sira Vegas}
\affiliation{%
	\institution{Escuela Técnica Superior de Ingenieros Informáticos, Universidad Politécnica de Madrid}
	\streetaddress{Campus Montegancedo, Calle ciruelos, s/n}
	\city{Boadilla del Monte, Madrid}
	\country{Spain}
	\postcode{28660}}
\email{svegas@fi.upm.es}


\begin{abstract}

Experiments are a commonly used method of research in software engineering (SE). Researchers report their experiments following detailed guidelines. However, researchers do not, in the field of test-driven development (TDD) at least, specify how they operationalized the response variables and, particularly, the measurement process. This article has three aims: (i) identify the response variable operationalization components in TDD experiments that study external quality; (ii) study their influence on the experimental results; (ii) determine if the experiment reports describe the measurement process components that have an impact on the results. Sequential mixed method. The first part of the research adopts a quantitative approach applying a statistical analysis of the impact of the operationalization components on the experimental results. The second part follows on with a qualitative approach applying a systematic mapping study (SMS). The test suites, intervention types and measurers have an influence on the measurements and results of the statistical analysis of TDD experiments in SE. The test suites have a major impact on both the measurements and the results of the experiments. The intervention type has less impact on the results than on the measurements. While the measurers have an impact on the measurements, this is not transferred to the experimental results. On the other hand, the results of our SMS confirm that TDD experiments do not usually report either the test suites, the test case generation method, or the details of how external quality was measured. A measurement protocol should be used to assure that the measurements made by different measurers are similar. It is necessary to report the test cases, the experimental task and the intervention type in order to be able to reproduce the measurements and statistical analyses, as well as to replicate experiments and build dependable families of experiments.
\end{abstract}


\received{9 August 2023}
\received[revised]{9 August 2023}
\received[accepted]{9 August 2023}

\maketitle

\section{Introduction}\label{sec:introduccion}

In order to be able to test an experimental hypothesis, it has to be operationalized~\cite{Bunge}. Operationalization is the process of connecting the conceptual definition of the hypothetical variables related to a specific set of measurement procedures~\cite{Neuman2002}. The procedures (and often the associated measurement instruments) depend on the area addressed. For example, test cases are usually used to measure the quality of code developed using TDD, whereas the quality of a specification could be determined using a code review checklist. The following measurement instruments are used in software engineering (SE): questionnaires, measurement software, expert evaluation, etc. In this study, we focused on the operationalization of the external quality response variable in TDD experiments. This is not fortuitous. We conducted a number of TDD experiments~\cite{Fucci2016, Dieste2017, santos2020increasing,santos2020family} and observed that the use of one or other procedure/instrument affects the external quality measurement \cite{dieste20}.

Strictly speaking, external quality refers to all the software product characteristics from an external viewpoint ~\cite{iso9126-1}. According to ISO/IEC 205010:2011~\cite{iso25010}, the software product external quality model is composed of characteristics and subcharacteristics. In TDD experiments, external quality refers to a combination of software product quality model completeness and functional correctness.

TDD experiments that study external quality produce different, and sometimes even contradictory, results. For example, studies by Geras et al. \cite{Geras2004} and George and Williams \cite{George2004} suggest that TDD improves quality. However, Huang and Holcombe \cite{Huang2009} and Desai et al. \cite{Desai2009} claim that the difference is very small and is not statistically significant. Madeyski \cite{madeyski2005preliminary} states that TDD reduces external quality. Researchers conduct secondary studies examining the experiment context (industry or academia) \cite{Causevic2011,Rafique} or the study’s rigour (research method carried out according to best practice) and relevance (real impact on industry ) \cite{Munir2014} in search of explanations for these inconsistencies, although the results are not absolutely satisfactory.

In our opinion, some of the differences in the results of TDD experiments that study external quality could be due to measurement-related issues \cite{dieste20}. The instruments usually used in TDD experiments to measure external quality are acceptance test cases \cite{Dieste2017, Santos2020}. External quality is proportional to the number of acceptance test cases passed \cite{Rafique2013}. However, many aspects related to the test cases are almost certain to vary across experiments, including, for example, the test case generation technique (ad hoc, equivalence partitioning, etc.), the person applying the test cases to the source code (subject, researcher, etc.), or the problem resolution criteria in the event of errors (method parameter inconsistency, etc.).

TDD analyses carried out by the authors of secondary studies (e.g., \cite{Bissi2016, Munir2014} have not taken into account these differences. In fact, the test cases used to measure the quality of the code generated by subjects could differ across studies, where the measurement process components and the impact of such differences are not well understood, and are, as a result, passed over by secondary studies. The same applies to other differences in external quality measurement.

The aim of our study is to identify the components of the response variable operationalization in TDD experiments that study external quality, investigate their influence on the experimental results, and determine whether the experimental reports describe the measurement process components that have an impact on the results in sufficient detail.

We used a sequential mixed method of research, which includes two procedures. The first adopts a quantitative approach based on statistically analysing the impact of the operationalization components on experimental results. The second follows on with a qualitative approach using a systematic mapping study (SMS), which tests and rounds out earlier findings on TDD experiment reporting.

The contributions of this article are as follows:

\begin{itemize}
	\item We gather empirical evidence on the impact of test cases and intervention type on the measurement of the value of external quality variable in TDD experiments. These effects could also occur in other SE experiments.
	
	\item We conclude that the measurer does not appear to have an impact on external quality measurement. This is rather surprising, as the human factor is generally considered to be an important source of noise in measurement.	
	
	\item We observe that the measurement process components are missing from TDD experimental reporting. The test cases, experimental task and intervention type need to be reported to be able to reproduce the measurements and analysis.
	
	\item Most TDD experimental reports explicitly define the research aims and discuss their results, but they do not provide the raw data, specify the derived data or attach the full replication protocol. Again, this information is necessary to assure experiment reproducibility, as well as to replicate experiments and build dependable families of experiments.

\end{itemize}

This paper is organized as follows. Section 2 describes the background and SE measurement process used in this paper. Section 3 describes the research method applied. Section 4 evaluates the impact of the measurement process components on the experimental results in TDD. Section 5 reports the result of the SMS. Section 6 defines the validity threats. Sections 7 and 8 discuss the results and the related work, respectively. Finally, Section 9 describes the conclusions and future work.

\section{Background}\label{sec:medición}

In this section, we describe the concept of operationalization. We then go on to briefly outline the FSecure experiment, which we used as a case study. Finally, we explain the measurement process applied in the above experiment following the guidelines proposed by Fenton and Bieman \cite{fenton2014software}.

\subsection{Operationalization}\label{sec:operacionalización}

A hypothesis on the cause-effect relationship represents the belief that there is a relationship between a cause construct and an effect construct~\cite{Wohlin}.  Scientific constructs often refer to imperceptible events~\cite{Bunge}.In order to test a hypothesis, it has to be operationalized. Operationalization is the process of connecting a conceptual definition with a specific set of measurement techniques or procedures~\cite{Neuman2002}. For example, the operationalization of the hypothesis:

\begin{center}
	{\itshape TDD improves external quality}
\end{center}
implies:
\begin{itemize}
	\item Deciding what TDD means, that is, test-first vs. test-first+refactoring, which will determine the treatment implemented in the experiment.
	\item Selecting a response variable that represents the ``external quality'' construct.
	\item Establishing a response variable measurement process.
\end{itemize}

In this study, we focus on the operationalization of the abovementioned external quality response variable. Different researchers may choose different response variables to represent ``external quality'' (e.g., test cases run, number of satisfied requirements, etc.) and different procedures for even measuring the same response variable (e.g., example, manual testing vs. automated testing). The experimental result should be the same irrespective of how it is measured. If the procedure affects the experimental results, the results are not valid~\cite{Wohlin}. This raises the question of \textbf{whether decisions made by researchers could have an impact on experimental results?} Similar arguments could likewise apply to the operationalization of the TDD treatment, and have, in fact, already been addressed as part of the study of TDD experiment conformance \cite{fucci2014impact}.

\subsection{The FSecure experiment}

In order to answer the above question, we use the FSecure experiment as a case study. We ran the FSecure (FS) experiment, described in \cite{Tosun2017}, together with other colleagues. The aim of the FS experiment was to compare the TDD and incremental test last (ITL) techniques with respect to software external quality and developer productivity in a software company setting from the viewpoint of researchers. Twenty-four professionals participated in the experiment conducted on FS premises in three different cities.

A pre-test/post-test design was used. All the subjects applied both treatments in a predetermined order. ITL and TDD were applied on the MarsRover API (MR) and Bowling Score Keeper (BSK) experimental tasks, respectively. Both MR and BSK are popular exercises used by the agile community. Java was used as the programming language and jUnit 4 as the testing framework. The subjects received the task specification and a source code template with a class structure and common methods to facilitate the measurement process.

\subsection{FS experiment measurement process}\label{sec:proceso-Fenton-Bieman}

Researchers usually use more specific processes than the abstract concept of operationalization. In SE, the process proposed by Fenton and Bieman \cite{fenton2014software} is probably the best known and is used as a guideline for describing the FS experiment measurement process.

\begin{figure}[h]
	\centering
	\includegraphics[width=\linewidth]{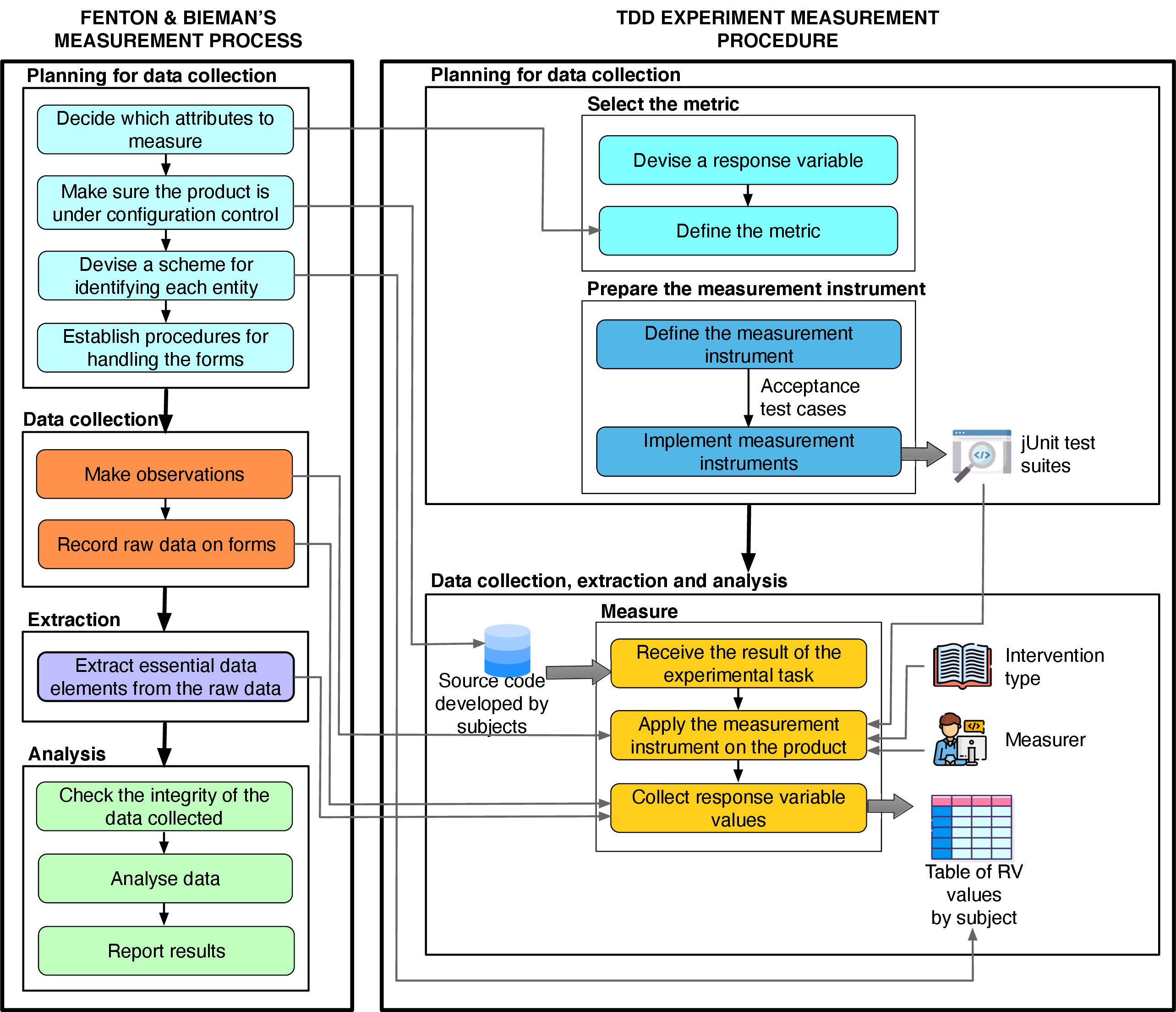}
	\caption{Measurement process according to Fenton and Bieman \cite{fenton2014software} vs. our measurement procedure}
	\label{fig:figura-procesos-medicion}
\end{figure}

According to Fenton and Bieman \cite{fenton2014software}, the measurement process is composed of four activities: (i) planning for data collection, (ii) data collection, (iii) extraction, and (iv) analysis. Each of these activities is further divided into several steps, which are shown on the left-hand side of Figure \ref{fig:figura-procesos-medicion}. They are mapped to the steps of the measurement procedure enacted in the FS experiment, illustrated on the right-hand side of Figure \ref{fig:figura-procesos-medicion}. This clearly depicts how many details have to be taken into account in a real measurement process.

\subsubsection{Planning for data collection}\label{sec:planning_DataCollection}

\textbf{Decide which attributes to measure:} This step is equivalent to the operationalization that we described in Section 2.3. It is extremely important at this point to clarify the definition of \textit{metric}. The terms \textit{measurement} and \textit{metric} are often used interchangeably in SE. The concepts related to measurement are defined in the \textit{Vocabulaire International de Métrologie}\footnote{\url{https://www.bipm.org/documents/20126/2071204/JCGM_200_2012.pdf/f0e1ad45-d337-bbeb-53a6-15fe649d0ff1}} (VIM). This standard defines \textit{measurement} is the process of experimentally obtaining one or more quantity values from a measurand. VIM does not include the term \textit{metric}, because this is a mathematical concept that expresses the distance between two points \cite[p. 120]{fenton2014software}. For the sake of precision, we use  the terms textit{measurement} and textit{measurement unit} hereinafter.

Two response variables were used in the FS experiment: external quality and productivity. In this study, we focus on external quality, as it is widely used in TDD experiments. However the results are equally applicable to productivity. In the FS experiment, we used the same variables and measurements as in \cite{Erdogmus2005,Fucci2013}, where external quality was defined as a combination of software product quality model completeness and functional correctness as defined in ISO/IEC 25010:2011~\cite{iso25010}. Measurement is made using acceptance test suites, and the measurement unit is the number of assertions passed.

Although we take acceptance test suites for granted, the fact is that, with the exception of experimental replications, different experiments use different test cases. \textbf{This is the primary reason for this research}. Test cases are not standard; they are designed by researchers. \textbf{The use of different test cases can lead to sizeable measurement differences~\cite{dieste20}}.

The FS experiment test cases are available at GitHub
\footnote{The test suites are provided as a single Eclipse workspace containing four projects. They are available as one Eclipse workspace at \url{https://github.com/GRISE-UPM/TestSuitesMeasurement/tree/master/test_suites}.}. One test suite was reused from previous experiments \cite{Erdogmus2005,Tosun2017}. It was generated using an \textit{ad-hoc} (AH) strategy and coded in jUnit. Ad hoc means that no formal procedure was used to create the test cases; the test suite authors (MR and BSK) applied their best judgement to write a set of test cases based on the functional requirements.

It is also necessary to specify which direct measurements are necessary and which measurements can be derived from the direct measurements. The external quality measurement is calculated using the number of tackled subtasks (\#tst) for a given task. A subtask is considered to have been tackled if it passes at least one assertion in the test case suite associated with the subtask. \#tst is calculated using Equation \ref{eq:tst}.

\begin{equation}
	\#tst = \sum_{i=0}^n \left\{ \begin{array}{rl}
		1 &\mbox{ if $ Num.Assert_{i} (PASS)>0$} \\
		0 &\mbox{ otherwise}
	\end{array} \right.
	\label{eq:tst}
\end{equation}

where \textit{n}, is the total number of subtasks that make up the experimental task. This measurement unit can differentiate between the subtasks that the experimental subject has and has not tried to complete. 

The quality of the \textit{i}-th subtask is calculated using Equation \ref{eq:qltyi} below:

\begin{equation}
	QLTY_{i} = \frac{\#Assert_{i} (PASS)} 
	{\#Assert_{i} (ALL)}
	\label{eq:qltyi}
\end{equation}

$\#Assert{_i} (PASS)$ represents the total number of assertions that pass the acceptance test cases associated with the \textit{i}-th subtask. $\#Assert{_i} (ALL)$ symbolizes the total number of assertions created to test the i-th subtask.

To calculate the quality of the experimental task implemented by the subject, we use the Equation below:

\begin{equation}
	QLTY = \frac{\sum_{i=1}^{\#tst}
		QLTY_{i}}  {\#tst}
	\label{eq:qlty}
\end{equation}

\noindent\textbf{Control configuration:}We should make sure that we know everything there is to know about the object to be measured. For example, it is common practice to use cross-over designs in TDD experiments. One property of these designs is that each experimental task is performed twice, once using TDD and again using the control treatment. Once the code has been stored, it is far from easy to determine whether or not the experimental task was resolved using TDD. We numbered treatments in order of execution. The code delivered by the participants in the FS experiment is available at GitHub\footnote{\url{https://github.com/GRISE-UPM/FiDiPro_ESEIL_TDD/tree/master/COMPANY_05}}.

\noindent\textbf{Devise a scheme for identifying each entity that is part of the measurement process:} It is important to clarify how the products, versions, installations, failures, etc., will be denoted on the data collection forms. In the FS experiment, each participant subject was given a code to assure anonymity. Each column is labelled to identify the measured object, and the measurement circumstances. In our case, the columns are source code identifier, experimental task, \#tst, quality measurement, session and treatment.

\noindent\textbf{Procedures to handle the forms, analyse the data and report results:} It has to be specified who fills in what, when and where, providing a clear description of how the completed forms will be processed. This step is secondary for our purposes.

\subsubsection{Data collection}
\noindent\textbf{Make observations:} This activity aims to obtain raw data. Raw data are the result of the preliminary measurement of the process, product or resource. The ideal thing is to capture data automatically. In many cases, however, there is no option but to collect data manually.

Raw data are the result of applying the acceptance test suites to the code delivered by the experimental subjects. However, the apparent simplicity of this step conceals \textbf{the other two fundamental problems motivating this research}:

\begin{enumerate}
	
	\item \textbf{How are conflicts resolved?} The test cases will most probably generate conflicts when applied to the code delivered by the experimental subjects. Common problems are method parameter addition and parameter type changes. However, there may be others, ranging from compilation errors to code incompatibility with test cases.
	
	We used three strategies in the FS experiment to connect acceptance test cases with the code delivered by subjects:
	
	\begin{itemize}
		\item No intervention: The code developed by subjects is measured as delivered, that is, the code is untouched.
		\item Syntactic intervention: The measurer makes changes to the code delivered by subjects to prevent minor errors during measurement.
		\item Semantic intervention: The measurer interprets the solution coded by the subject and makes major changes to the delivered source code.
	\end{itemize}
	
	\item \textbf{Who resolves the conflicts?} Conflict resolution cannot generally be performed automatically. A measurer must resolve the conflicts, and this may affect measurement.
\end{enumerate}

In other words, \textbf{we believe that there is a profound gap between the operationalization and measurement of the variable, which SE experimentation has overlooked until now}. The measurement instruments used are odd, at least in TDD and possibly in other SE areas. Instead of measurement rules or instruments related to well-defined units of measurement, like forces or electric charges, SE uses units of measurement, like acceptance tests passed, which are unique and whose relationships to the measurement instruments are largely ad hoc. We also believe, as illustrated in Section~\ref{sec:eval}, that these instruments and their application have to be taken into account.


\noindent\textbf{Record raw data on forms:} Either automatically (preferably) or manually. In our case, the measurements were stored manually in Excel files, available at Google Sites\footnote{\url{https://sites.google.com/site/fidiproeseil/home}}. The possibility of the measurer entering incorrect data cannot be ruled out, which means that transcription is another source of error that may be amplified by the measurer.

\subsubsection{Extraction}
In Step 3, the \textbf{essential data elements are extracted from the raw data} so that the analysts can derive attribute values. In the FS experiment, Excel formulas were created to implement Equation~\ref{eq:qlty}, and QLTY was measured directly from the raw data. This ruled out errors of calculation.


\subsubsection{Analysis}
Finally, it is necessary to \textbf{check data integrity, and analyse and report the results}. These activities are secondary to our purposes and are not addressed in depth here.

\section{Objectives and methodology}\label{sec:metodologia}

To the best of our knowledge, the impact of the measurement procedure on experimental results has not yet been studied in SE and definitely not in TDD. Therefore, we planned our research as a sequential mixed method, composed of a quantitative phase, followed by a qualitative procedure \cite{creswell2017research}. The quantitative research question is:

\textbf{RQ-QUAN: What impact do test cases, intervention type and measurers have on the external quality of TDD experiments?}

The quantitative procedure includes a series of analyses using the Bland-Altman plot~\cite{Bland1986} applied to the published dataset of the FS experiment conducted in industry \cite{Tosun2017} to compare measurements and check the impact of the three elements that are overlooked in the TDD external quality measurement process: (i) test case design, (ii) intervention type, and (iii) measurer. This quantitative phase of the mixed design is detailed in Section~\ref{sec:eval}. 

The general qualitative research question is:

\textbf{(RQ-QUAL): What measurement process elements are reported in the experiments on TDD that study external quality?}

After identifying the operationalization elements that have an impact on response variable measurement, a SMS, conducted following guidelines by Kitchenham and Charters \cite{keele2007guidelines}, is used to check whether the respective elements are reported in the TDD experiments published in the literature. This qualitative phase of the mixed design is detailed in Section 5. 

The research question that integrates both the quantitative and qualitative procedures is:

\textbf{(RQ-QUAN+QUAL): How does TDD experiment reporting facilitate knowledge creation?}

The objective is to determine what impact the operationalization process elements have on the measurement process with respect to replication and the formation of families of experiments. This question is discussed in Section \ref{sec:discusion} to output some recommendations for improving the response variable measurement process and, consequently, the results of the experiments conducted in TDD and SE as a whole.

\section{Evaluation of the impact of the measurement process on experimental results (RQ-QUAN)}\label{sec:eval}

As mentioned in Section~\ref{sec:medición}, a number of activities have to be performed to enact the measurement process applied to TDD experiments that evaluate external quality. They are:

\begin{itemize}
	\item Design and code the test suites
	\item Connect the code delivered by subject with the test suites, resolving syntactic and semantic inconsistencies 
	\item Collect the pass/failure information for the test cases.
\end{itemize}

These activities could influence the results of the measurement. Indeed, Dieste et al.~\cite{dieste20} already found that the use of different test cases may have a major impact on the response variable values. As a result, the associated statistical analyses may be significant in one case and not significant in another or the effects may be positive in one case and negative in another. This article replicates \cite{dieste20} and extends their findings.

In the following, we quantitatively analyse the influence of the above activities using the dataset of the FS experiment conducted in industry \cite{Tosun2017}. The main test method that we use is the Bland-Altman plot \cite{Bland1986}. Although there are other measurement comparison methods, like ISO 5725 \cite{ISO5725}, that provide more synthetic results, the Bland-Altman plot is a graphical technique that helps to visualize the differences between measurements. For an overview of the different methods of comparison available, see Dieste et al.~\cite{dieste20}.

\subsection{Test cases}\label{sec:testcases}

\subsubsection{Description of test cases}\label{subsec:experimentoFS}

TDD experiments that study external quality often use test cases \cite{Canfora2006, Causevic2012, Desai2009, Fucci2016} \footnote{Other studies measure external quality considering the number of defects detected by bug tracking systems \cite{Bhat2006, Nagappan2008}. However, these defects may have been identified using a wide range of procedures. Therefore, we focus on test cases.}. The test cases are run on the source code delivered by subjects. The formulas described in Section \ref{subsec:impacto} are applied to the test cases run to yield the measurement value.

In the FS experiment \cite{Tosun2017}, we relied on an acceptance test case suite for each task (MR and BSK). These test case suites were generated using two different techniques: ad hoc (AH) and equivalence partitioning (EP), respectively \footnote{The test suites are provided as a single Eclipse workspace containing four projects. They are available as one Eclipse workspace at \url{https://github.com/GRISE-UPM/TestSuitesMeasurement/tree/master/test_suites}.}.

No formal procedure was used to generate the AH test cases. The term \textbf{ad hoc} refers to a specific solution developed for a particular problem. The authors of test case suites used their best judgement to create the test case suite.  

The EP technique was also used to develop the test case suite. EP is based on the fact that component inputs and outputs can be divided into equivalence classes according to the specifications. There are two possible types of equivalence classes: valid equivalence classes that represent valid inputs for the program and invalid equivalence classes that represent all the other possible states of the condition \cite{myers2011art}. The result of testing a single value of the equivalence partition is considered to be representative of the whole partition \cite{bcs2001}.

Table \ref{tab:AH_vs_EP} shows the number of test classes/methods/assertions of each suite for the MR and BSK tasks. All the test case suites were coded in the jUnit test framework.

The result for external quality (QLTY) output by the test suites is a percentage (0 \%-100 \%). This percentage represents, for example, the extent to which the code complies with the software requirements: a value of 0 \% means that the code does not satisfy any requirement, and a value of 100 \% means that the code meets all the requirements.

\begin{table}
	\centering
	\small
	\caption{Characteristics of AH and EP test suites}
	\label{tab:AH_vs_EP}
	\begin{tabular}{llcc}
		\toprule
		& \textbf{} & \multicolumn{2}{c}{\textbf{Test suite}} \\ \cline{3-4} 
		Task &  & AH & EP \\ \cline{1-4} 
		
		\multirow{3}{*}{MR} & Test classes & 11 & 9 \\
		& Test methods & 52 & 32 \\
		& Assertions & 89 & 32 \\ \cline{1-4} 
		\multirow{3}{*}{BSK} & Test classes & 13 & 13 \\
		& Test methods & 48 & 72 \\
		& Assertions & 55 & 72 \\ \cline{1-4} 
		
	\end{tabular}
\end{table}

Dieste et al. \cite{dieste20} analysed the test case suites and found that AH and EP are more or less equivalent. The statement coverage yielded very similar results (100\%) for both AH and EP. Branch coverage also output similar results (90\%), except for the AH test case suite applied to the MR task, which yielded slightly lower outcomes (88\%). These coverages suggest that the results of testing using both test suites will be similar. For example, a simple correlation analysis to evaluate convergent validity returns large, statistically significant correlations ($ r> 0.5 $, according to Cohen \cite{Cohen1988}).

\subsubsection{Impact on statistical analyses}\label{subsec:impacto}

The FS experiment was designed as a pre-test/post-test experiment \cite{shadish2002experimental}. The applicable data analysis for this type of experiment is

\begin{equation}
	\label{eq:FS-analysis}
	QLTY = Treatment + \epsilon
\end{equation}

\begin{figure}[h]
	\center
	\includegraphics[width=0.5\linewidth]{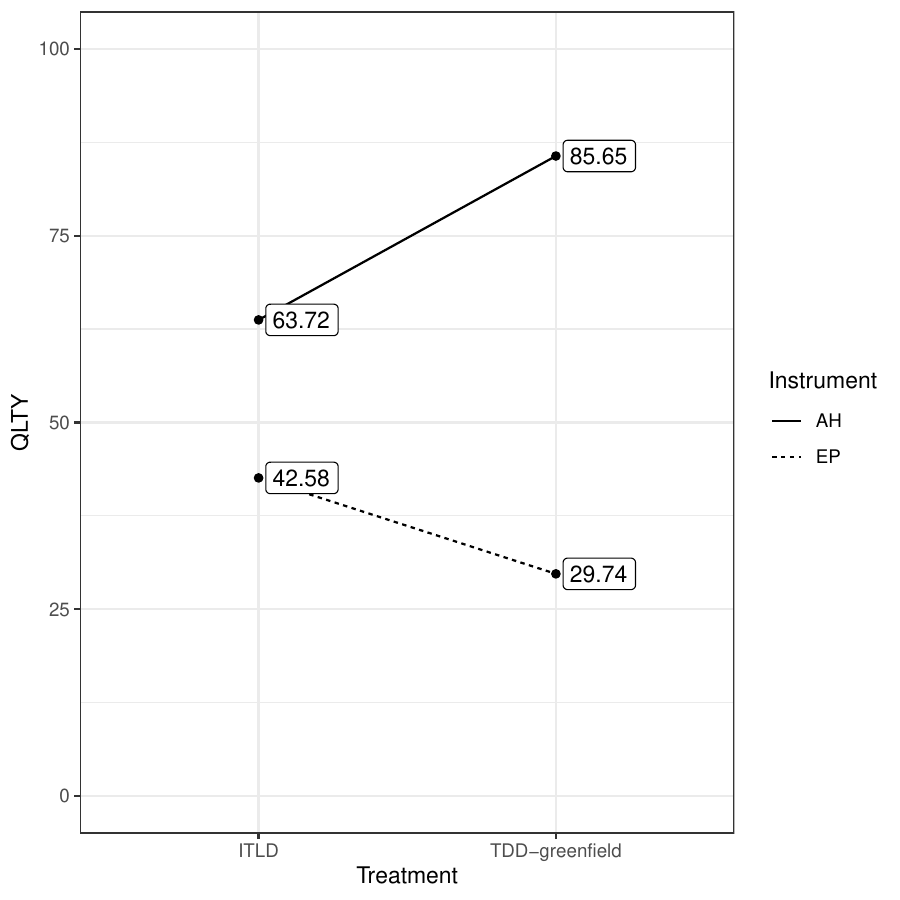}
	\caption{Effect of ITL/TDD depending on whether the measurement is made using the AH or EP test cases}
	\label{fig:AHvsEP-graphycally}
\end{figure}

Table \ref{tab:fernando-testcases-equivpart-syntactic} shows the analysis of external quality measured using the AH test suites (originally reported in \cite{Tosun2017}) and EP. There are clear differences. The intercept is similar (which is to be expected for positive and bounded response variables). However, the effect of TDD is positive for the AH suite (21.93 points) and negative for EP (-12.48 points), and the results are statistically significant in both cases. The profile plot in Figure \ref{fig:AHvsEP-graphycally} illustrates the information reported in Table \ref{tab:fernando-testcases-equivpart-syntactic} more clearly. The solid and dashed lines represent the effect of TDD as measured by AH and EP, respectively. It is clear that AH and EP yield opposite results, especially for TDD.

\begin{table}
	\small
	\centering
	\caption{Statistical analysis of QLTY response variable for AH and EP test cases}
	\label{tab:fernando-testcases-equivpart-syntactic}
	\begin{tabular}{l c c}
		\hline
		& Ad-hoc & Equivalence partitioning \\
		\hline
		(Intercept)             & $63.72 \; (6.00)^{***}$ & $42.58 \; (4.50)^{***}$ \\
		TreatmentTDD-greenfield & $21.93 \; (8.48)^{*}$   & $-12.84 \; (6.37)^{*}$  \\
		\hline
		R$^2$                   & $.13$                   & $.08$                   \\
		Adj. R$^2$              & $.11$                   & $.06$                   \\
		Num. obs.               & $48$                    & $48$                    \\
		\hline
		\multicolumn{3}{l}{\scriptsize{$^{***}p<0.001$; $^{**}p<0.01$; $^{*}p<0.05$}}
\end{tabular}

\end{table}

\subsubsection{Source of the differences in statistical analyses}

There is only one possible cause behind the differences in the results of the statistical analyses: there are differences in the values of the \textit{QLTY} response variable for AH and EP. The Bland-Alman plot is a mechanism for visualizing such differences \cite{Bland1986}.

The Bland-Altman method \cite{Bland1986} is the de facto standard for comparing measurement instruments in medicine, but it can be easily adopted in SE. In our research problem, QLTY has been measured twice using different instruments: the AH and EP test suites. We thus have two sets of measures $Y_{AH} = \{y_{AH_1}, y_{AH_3}, \dots, y_{AH_n}\}$ and $Y_{EP} = \{y_{EP_1}, y_{EP_3}, \dots, y_{EP_n}\}$, where $1 \leq i \leq n$ are the different pieces of code measured. 

The Bland-Altman method starts with the calculation of the difference between measurements:

\begin{equation}
	\label{eq:bland-altman-difference}
	d_i=(y_{AH_i} - y_{EP_i})
\end{equation}

Next, the average of the differences $d_i$ is calculated as:

\begin{equation}
	\label{eq:bland-altman-mean-difference}
	\bar{d}=\frac{1}{n}\sum\limits_{i = 1}^n {(y_{AH_i} - y_{EP_i})}
\end{equation}

$\bar{d}$ represents the mean difference between the measurements obtained with the AH and EP test suites. Finally, the standard deviation is calculated as:

\begin{equation}
	\label{eq:bland-altman-sd-difference}
	s_d=\sqrt{\frac{1}{n-1}\sum\limits_{i = 1}^n {(d_i - \bar{d})^2}}
\end{equation}

\begin{figure}[h]
	\centering
	\includegraphics[width=0.5\linewidth]{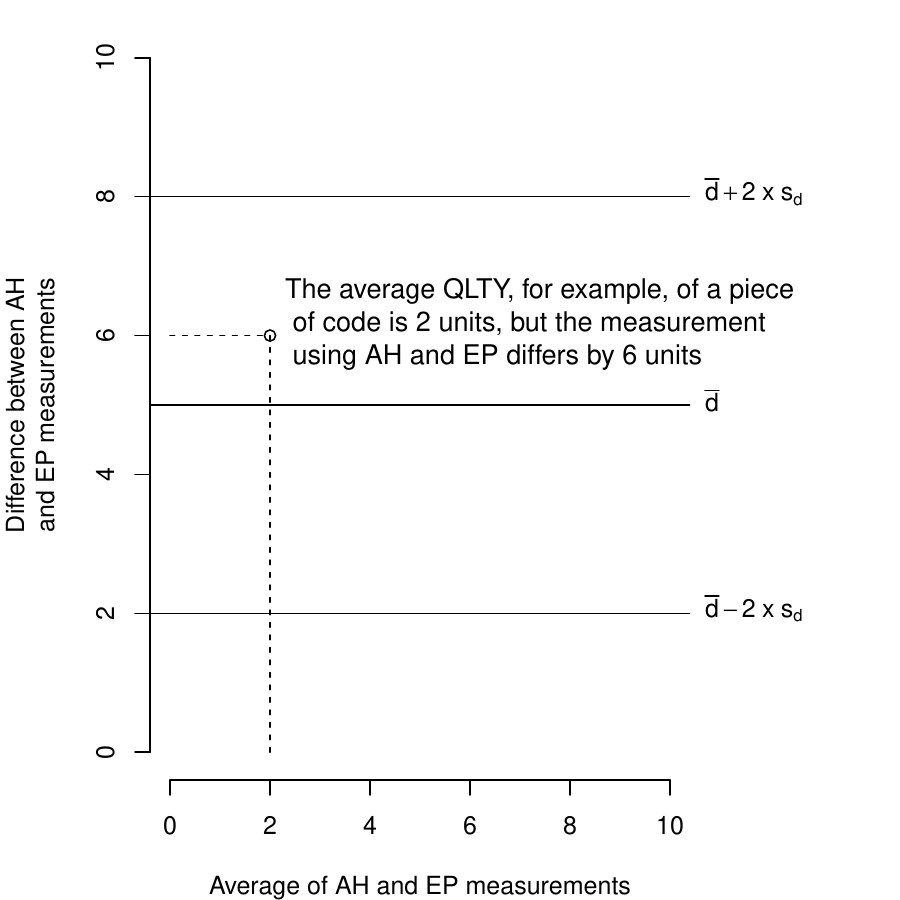}
	\caption{Difference between AH and EP measurements}
	\label{fig:example-bland-altman}
\end{figure}

The Bland-Altman method has an associated graphical representation (the Bland-Altman plot), shown in Figure~\ref{fig:example-bland-altman}. This graph plots the mean values obtained by both measurement instruments:

\begin{equation}
	\label{eq:bland-altman-mean}
	\frac{y_{AH_i} + y_{EP_i}}{2}
\end{equation}

that is, the best estimation of the true measurements against their difference:

\begin{equation}
	\label{eq:bland-altman-difference}
	y_{AH_i} - y_{EP_i}
\end{equation}

A horizontal line is drawn at the mean difference $\bar{d}$. Additionally, the graph also depicts two additional horizontal lines located at

\begin{equation}
	\label{eq:bland-altman-limits}
	\bar{d} \pm 2 \times s_d
\end{equation}

Assuming a normal distribution for the differences, these bounds enclose 95\% of the differences $d_i$. \textbf{These bounds represent the variability in the measurement of the same code when one (AH) or other (EP) text suite is used}.

The Bland-Altman plots shown in Figure~\ref{fig:bland-altman-plots-fernando-equivpart-tasks} illustrate the distribution of the differences between the measurements made using the AH and EP test cases. In the FS experiment, MR was used as the control task in the first phase of the experiment (ITL trial), and BSK was employed in the second phase (during TDD use). As the treatment and task are confounded, separate Bland-Altman plots need to be created for the two tasks in order to analyse the differences that highlighted in Table~\ref{tab:fernando-testcases-equivpart-syntactic}.

Figure~\ref{fig:bland-altman-plots-fernando-equivpart-tasks} brings to light the following characteristics:

\begin{itemize}
	
	\item The differences (AH-EP) are generally greater than 0. The measurement values with AH are greater than the measurements using EP.
	
	\item On average, the measurements made with AH are 21.14 points higher than measurements using EP for the MR task. In the case of the BSK task, the differences are much larger, with a difference of 59.91 points.
	
	\item The measurements fluctuate more for MR than for BSK. Figure~\ref{fig:bland-altman-plots-fernando-equivpart-MR} shows that the measurements can vary by $\pm 32.98$ points with respect to the average differences. For BSK (see Figure~\ref{fig:bland-altman-plots-fernando-equivpart-BSK}), the differences are much smaller ($\pm 17.64$ points).

\end{itemize}

There is no simple explanation for why the AH and EP test suites provide such different measurements. For a clarification of the causes, see Dieste at al.~\cite{dieste20}. This research is concerned not with the reasons for the above differences, but with the fact that they exist and what impact they have on the experimental results, as discussed below.

\subsubsection{Impact of the differences between measurements on statistical analyses}

\begin{figure}[h]
	\begin{subfigure}{6cm}
		\centering\includegraphics[width=6cm]{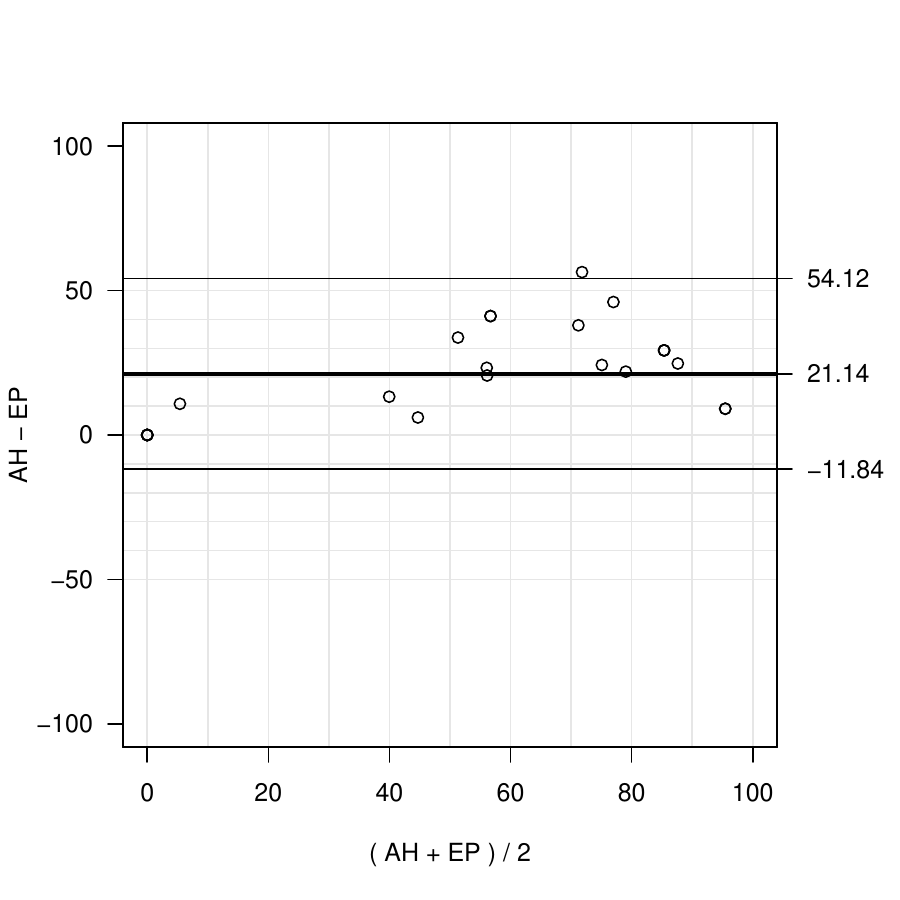}
		\caption{MR}
		\label{fig:bland-altman-plots-fernando-equivpart-MR}
	\end{subfigure}
	\begin{subfigure}{6cm}
		\centering\includegraphics[width=6cm]{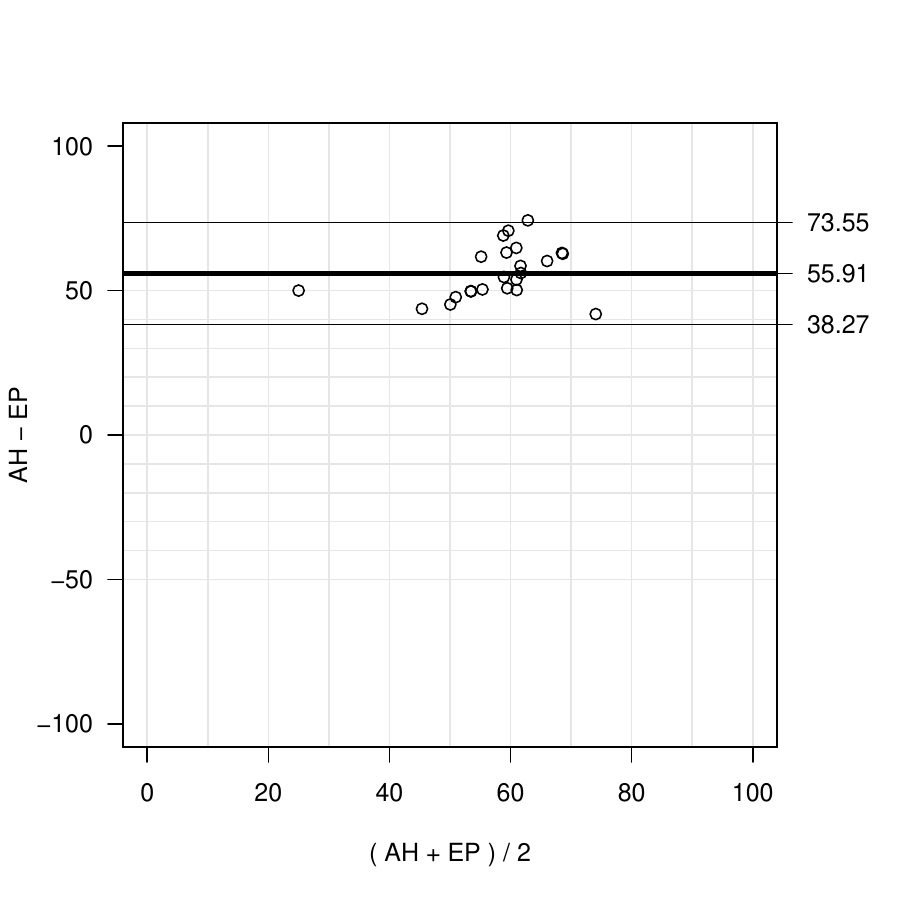}
		\caption{BSK}
		\label{fig:bland-altman-plots-fernando-equivpart-BSK}
	\end{subfigure}
\caption{Bland-Altman plots for differences between AH and EP test cases.}
\label{fig:bland-altman-plots-fernando-equivpart-tasks}
\end{figure}

The differences in the measurements made using the AH and EP test cases are a possible explanation for the deviations in the statistical analyses mentioned above:

\begin{itemize}
	
	\item The average differences between measurements explain the change in the treatment effect in Table~\ref{tab:fernando-testcases-equivpart-syntactic}. Note that AH produces higher measurements than EP, and these differences are greater for the BSK task. As TDD and BSK are confounded, AH is favourable to TDD, which scores $55.91-21.14=34.77$ points more than ITL/MR. As Table~\ref{tab:fernando-testcases-equivpart-syntactic} shows, $-12.84+34.77=21.93$.
	
	\item As measurements made with AH are more variable, its standard deviation is greater than for EP. Although it is not as easy to numerically relate Table~\ref{tab:fernando-testcases-equivpart-syntactic} and Figure~\ref{fig:bland-altman-plots-fernando-equivpart-MR}, in this case, we can see that the standard deviation of the AH measurement ($8.48$) is greater than for EP ($6.37$). In this case, the larger variance does not have an effect on the analysis, but it could possibly transform a statistically significant into a statistically non-significant result.
	
\end{itemize}

\subsection{Intervention types}\label{sec:interventions}

\subsubsection{Description of intervention types}\label{subsec:describe_tipos}

The test cases and the code to be measured can be connected in different ways, which we refer to as \textit{intervention types}. Although there are many possible variants, our experience from the ESEIL project \cite{juristo2016experiences} suggests that there are three possible ways of manipulating code for measurement using test cases: (i) no intervention, (ii) syntactic intervention, and (iii) semantic intervention. We describe each intervention type below.

\textit{No intervention}: In this type of measurement, the code delivered by the subjects is measured as is without any intervention at all by the measurer. Any minor errors in the source code (for example, a missing semicolon) are not corrected. 

\textit{Syntactic intervention}: The measurer makes some changes to fix minor bugs in the source code delivered by the subjects. These errors are easily identifiable in a debugging session, and the measurement values for code containing uncorrected errors would be lower than warranted, as the code containing the minor bug would fail the test cases. The main operations for carrying out a syntactic intervention are as follows:

\begin{itemize}
	\item Rename the classes of the source code delivered by subjects using the test case labels.
	\item Rename the methods of the code delivered by subjects using test case labels.
	\item Correctly assign the variables that return the results of a method.
	\item Modify the output format of the methods to what is expected by the test cases.
	\item Create the constructors with the parameters used in the test cases.
	\item Create the methods used in the test cases in the code delivered by the subjects.
	\item Assign suggested initial values to the variables, for example: \textit{xStart=0}.
\end{itemize}

Figure \ref{fig:Sintactica1} shows an example of source code delivered by an experimental subject. The code represents two functions that return a value (\textit{public int getfirstThrowValue()} and \textit{public int getSecondThrowValue()}). However, these function names are incorrect insofar as they are not the function names used in the measurement test cases. The subject does not appear to have followed the suggested source code template.

\begin{figure} [h]
	\small
	\begin{lstlisting}[language=Java]
		public int getfirstThrowValue() {
			
			return this.firstThrow;
		}
		
		public int getSecondThrowValue() {
			
			return this.secondThrow;
		}
		
	\end{lstlisting}
	\normalsize
	\caption{Example of source code delivered by an experimental subject for the BSK task}
	\label{fig:Sintactica1}
\end{figure}

Figure \ref{fig:Sintactica2} shows the required syntactic intervention on the source code: the measurer changes the name of the functions to the names used in the measurement test cases (\textit{public int getThrow1 ()} and \textit{public int getThrow2()}). Without this intervention, the test cases would not connect to the code delivered the subject, and the measurement value would not be correct.

\begin{figure} [h]
	
	\small
	\begin{lstlisting}[language=Java]
		public int getThrow1() {
			
			return this.firstThrow;
		}
		
		public int getThrow2() {
			
			return this.secondThrow;
		}
		
		
	\end{lstlisting}
	\normalsize
	\caption{Example of syntactic intervention on the source code delivered by the subject}
	\label{fig:Sintactica2}
\end{figure}

\textit{Semantic intervention}: As the measurement of the FS experiment progressed, we found that the source code needed to be analysed in more detail in order to better understand how the experimental subject solved the experimental task. In this respect, minor changes are not always enough, and substantial interventions are sometimes required to adapt the code delivered by the subject to the test case class structure and methods. The aim is to get a measurement that evaluates  the code implemented by the subject, which does not necessarily adhere to the solution that the researchers had in mind. Developers can come up with surprisingly imaginative and original solutions. It would be unfair to regard such code as invalid just because it does not adopt the expected solution. We refer to these changes as \textit{semantic} intervention, because measurers have to understand the code delivered by the subjects, which then has to be carefully connected to the test cases before making the measurement.

In the context of the BSK experimental task, bonus is an extra score added to the bowling game scoreboard. Figure \ref{fig:BSKbonus1} shows the source code delivered by an experimental subject developing BSK. The subject uses two methods to calculate the score: \textit{setBonus(int firstThrow, int secondThrow)} and the \textit{score()} function. In this second option, the bonus is calculated using the bonus variable, which is correct.\\

\begin{figure} [h]
	
	\scriptsize
	\begin{lstlisting}[language=Java]
		public void setBonus(int firstThrow, int secondThrow) {
			bonus = new Frame(firstThrow, secondThrow);
		}
		
		public int score(){
			int gameScore = 0;
			for(int i = 0; i< frames.size();i++){
				Frame f = frames.get(i);
				int frameScore = f.score();
				if(f.isStrike()){
					if(f.isLastFrame()){
						if (bonus!=null)
						frameScore+=bonus.score();
					}else{
						Frame nf = frames.get(i+1);
						frameScore += nf.score();
						if(nf.isStrike()){
							if(nf.isLastFrame()){
								if (bonus!=null)
								frameScore+=bonus.getFirst(); 
							}else{
								frameScore+=frames.get(i+2).getFirst();
							}
						}
						
					}
				}else if(f.isSpare()){
					if(f.isLastFrame()){
						if (bonus != null)
						frameScore+=bonus.getFirst();
					}else{
						Frame nf = frames.get(i+1);
						frameScore+= nf.getFirst();
					}
				}
				gameScore+=frameScore;
			}
			return gameScore;
		}
	\end{lstlisting}
	\normalsize
	
	\caption{Implementation of the score calculation in the BSK task using a bonus variable}
	\label{fig:BSKbonus1}
\end{figure}

However, this is not the only solution for calculating the BSK score. As shown in Figure \ref{fig:BSKbonus2}, subjects sometimes develop other types of implementations. In this case, the score is calculated considering the bonus within the frame array instead of the bonus variable as above.

\begin{figure} [h]
	
	\small
	\begin{lstlisting}[language=Java]
		public void setBonus(int firstThrow, int secondThrow) {
			//to be implemented
		}
		
		public int score(){
			int result = 0; 
			for (Frame f : frames) {
				if (!f.isBonus()) {
					result += f.score(); 
				}
			}
			return result; 
		}
	\end{lstlisting}
	\normalsize
	\caption{Alternative implementation of bonus in the BSK task}
	\label{fig:BSKbonus2}
\end{figure}

This second option is also valid. In semantic intervention, the measurers identify valid solutions like the above. The activities performed by measurers in order to understand the source code delivered by subjects and make any modifications that are necessary for connection to the test cases are as follows.

\begin{itemize}
	\item Review the classes of the source code delivered by the subject.
	\item Review each of the test cases that fail.
	\item Review the errors with the help of the development environment debugger.
\end{itemize}

To check the impact of the intervention type, the original code was measured by F. Uyaguari using the three intervention types and by D. Fucci also using the three intervention types. The AH test case suite was used to make the measurements. These six measurements are new and exclusive to this study, that is, we do not use the measurements reported in Tosun et al. \cite{Tosun2017}.

\subsubsection{Impact on statistical analyses}

Table~\ref{tab:fernando-intervention} shows the results of the analyses of the \textit{ITL} and \textit{TDD-greenfield} treatments depending on the different intervention types. In this case, the measurements were made by F. Uyaguari. The measurements made by D. Fucci have the same characteristics as the measurements made by F. Uyaguari, as shown in Section~\ref{sec:davide} (see Table~\ref{tab:davide-intervention}).

The variations in the results of the analyses are less pronounced than for the test cases, but they are there all the same.

\begin{itemize}
	
	\item The effect size differences between \textit{ITL} and \textit{TDD-greenfield} drop as the intervention type becomes less rigorous. There is quite a substantial variation: the effect size of \textit{TDD-greenfield} for the \textit{no intervention} type is $49.52$ points, which drops to $16.15$ for the \textit{semantic intervention} type.
	
	\item The change also affects the statistical significance, that is, the p-value increases for the \textit{syntactic} and \textit{semantic intervention} types.
	
\end{itemize}

This paper, focusing exclusively on measurement, is not the place to discuss which type of intervention is best. Note, however, that, as Table \ref{tab:fernando-intervention} shows, the experimental results may vary substantially depending on the intervention type used.

\begin{table}[h]
	\small
	\centering
	\caption{Statistical analysis of the QLTY response variable for the three intervention types (measurer: F. Uyaguari)}
	\label{tab:fernando-intervention}
	\begin{tabular}{l c c c}
		\hline
		& No Intervention & Syntactic & Semantic \\
		\hline
		(Intercept)             & $27.57 \; (6.70)^{***}$ & $63.72 \; (6.00)^{***}$ & $73.35 \; (4.75)^{***}$ \\
		TreatmentTDD-greenfield & $49.52 \; (9.47)^{***}$ & $21.93 \; (8.48)^{*}$   & $16.15 \; (6.72)^{*}$   \\
		\hline
		R$^2$                   & $.37$                   & $.13$                   & $.11$                   \\
		Adj. R$^2$              & $.36$                   & $.11$                   & $.09$                   \\
		Num. obs.               & $48$                    & $48$                    & $48$                    \\
		\hline
		\multicolumn{4}{l}{\scriptsize{$^{***}p<0.001$; $^{**}p<0.01$; $^{*}p<0.05$}}
\end{tabular}\end{table}

\subsubsection{Sources of the differences in the statistical analyses}\label{subsec:source_diff_stat}

\begin{figure}[!htbp]
	\begin{subfigure}{6cm}
		\centering\includegraphics[width=5.5cm]{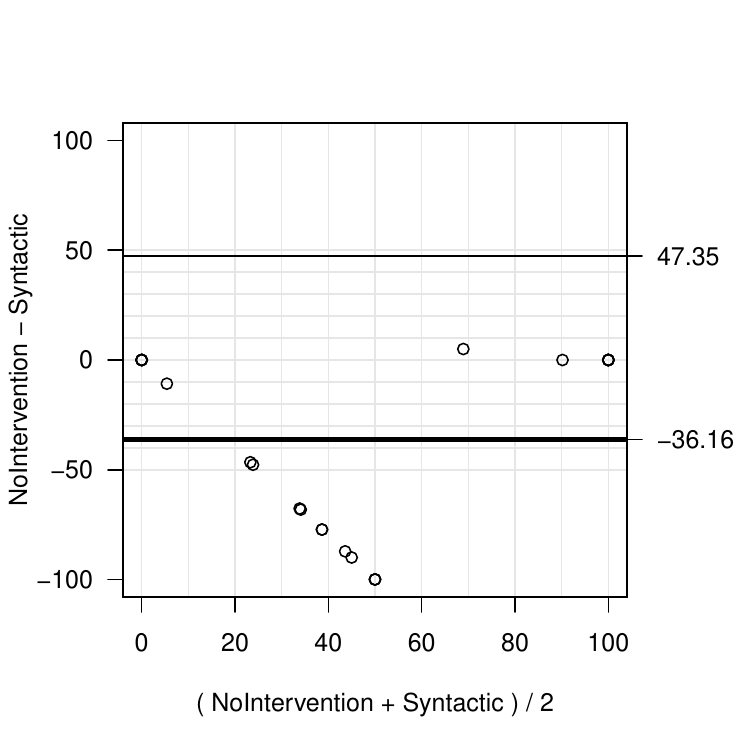}
		\caption{MR}
	\end{subfigure}
	\begin{subfigure}{6cm}
		\centering\includegraphics[width=5.5cm]{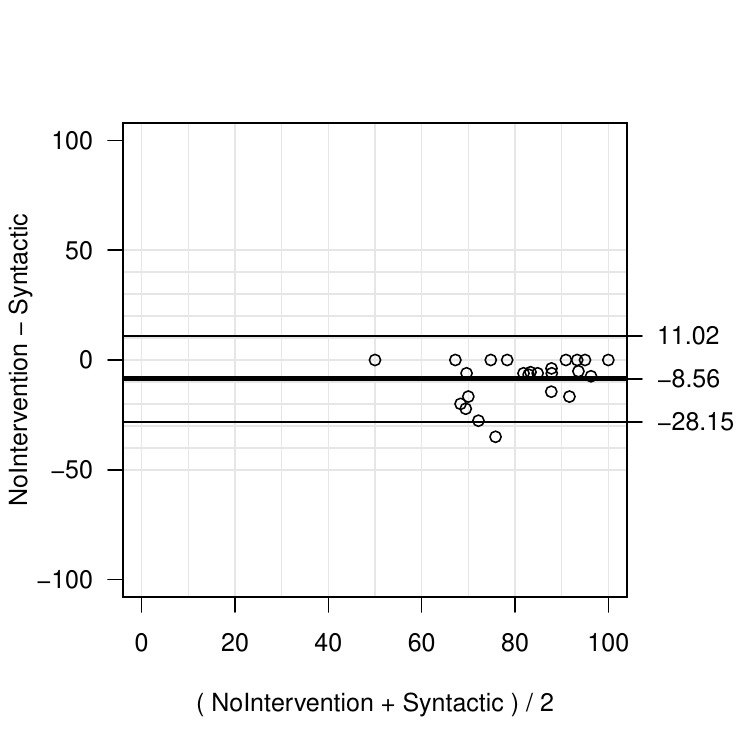}
		\caption{BSK}
	\end{subfigure}

	\begin{subfigure}{6cm}
		\centering\includegraphics[width=5.5cm]{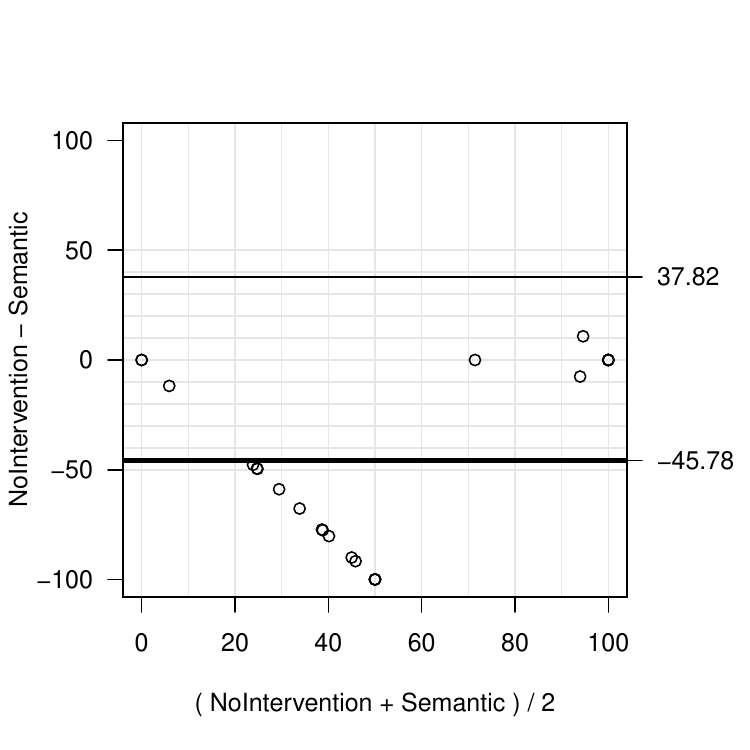}
		\caption{MR}
	\end{subfigure}
	\begin{subfigure}{6cm}
		\centering\includegraphics[width=5.5cm]{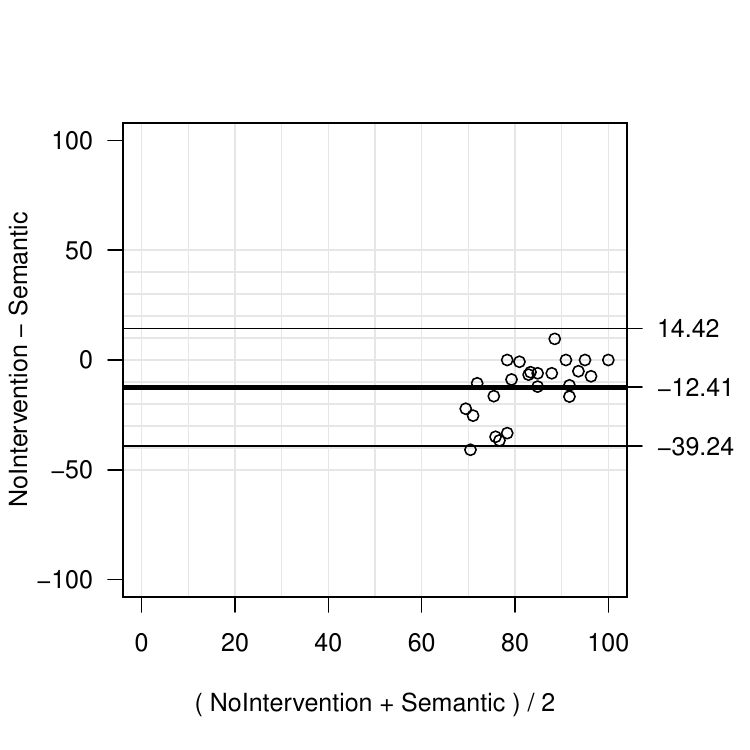}
		\caption{BSK}
	\end{subfigure}

	\begin{subfigure}{6cm}
		\centering\includegraphics[width=5.5cm]{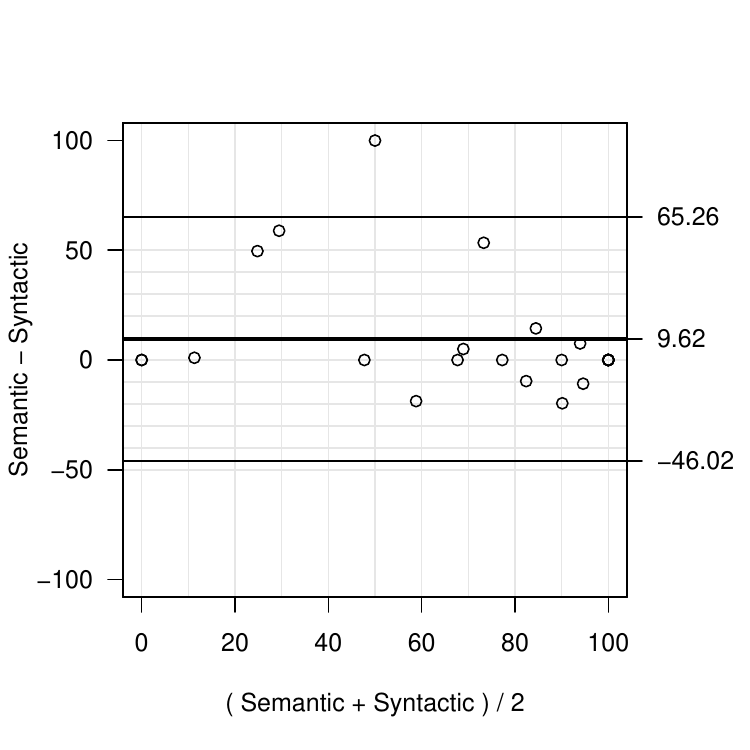}
		\caption{MR}
	\end{subfigure}	
	\begin{subfigure}{6cm}
		\centering\includegraphics[width=5.5cm]{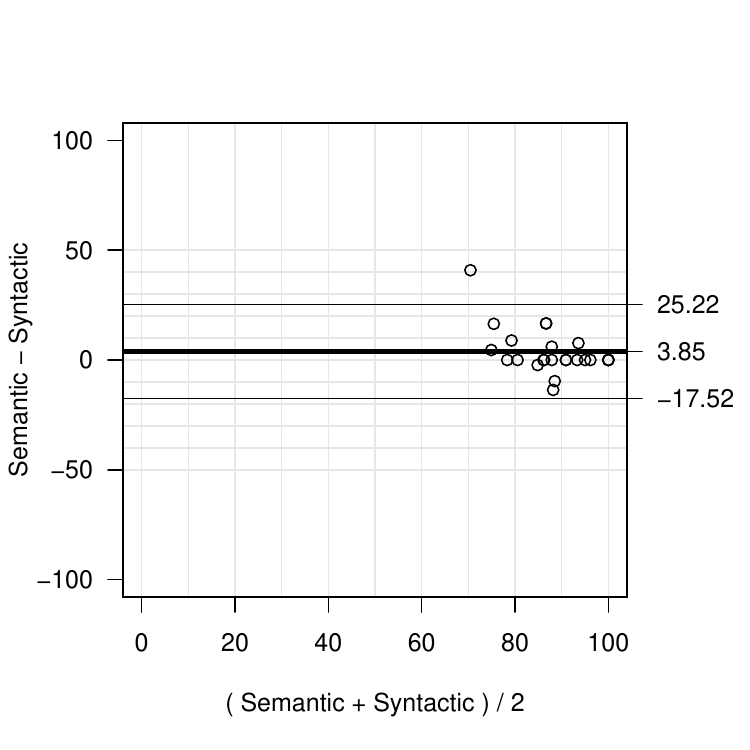}
		\caption{BSK}
	\end{subfigure}

	\caption{Bland-Altman plots for differences between intervention types (measurer: F. Uyaguari).}
	\label{fig:bland-altman-plots-fernando-intervention}
\end{figure}

The three possible Bland-Altman plots for the three pairs of intervention types are shown in Figure \ref{fig:bland-altman-plots-fernando-intervention}, differentiated, as in Section~\ref{sec:testcases}, by experimental task (MR or BSK). The differences in the measurements evidenced by the above plots are as follows:

\begin{itemize}
	
	\item As before, the biggest differences between the measurements are for the MR task. For BSK, the differences are much smaller.
	
	\item The bias or average difference between the measurements made with the \textit{no Intervention} type and either of the other \textit{syntactic}/\textit{semantic intervention} types is quite large. Predictably, the difference between \textit{no intervention} and \textit{semantic intervention} is greater than the difference between \textit{no intervention} and \textit{syntactic intervention}, as the differences are proportional to the level of source code manipulation.
	
	\item The bias between the measurements after \textit{semantic intervention} and \textit{syntactic intervention} is relatively small for both MR and BSK.
	
	\item In any of the three cases, the confidence interval bounds of the difference between the measurements are very wide. This appears to be due to the presence of extreme differences, which, in several cases, can be regarded as outliers and are clearly visible in the Bland-Altman plots shown in Figure \ref{fig:bland-altman-plots-fernando-intervention}. MR is more susceptible than BSK to this problem.
	
\end{itemize}

\subsubsection{Impact of the differences on the statistical analyses}

The differences between the measurements obtained with the different intervention types —no intervention, syntactic intervention and semantic intervention— can be easily mapped to the results of the statistical analyses in Table~\ref{tab:fernando-intervention}. The reasoning is similar to the arguments reported in Section~\ref{subsec:source_diff_stat}.

\begin{itemize}
	
	\item The average differences between MR and BSK cause the \textit{treatment} to have a smaller effect. For example, comparing no intervention and syntactic intervention, there is a difference of $-36.16-(-8.56)=-27.6$ points, which corresponds to the difference between the effects $21.93$ and $49.52$, shown in Table~\ref{tab:fernando-intervention}.
	
	\item The bigger fluctuations in the MR measurements for the \textit{no intervention} and \textit{syntactic intervention} types increase the standard deviation (and could quite possibly reduce the significance of the results). 
	
	\item However, the p-values would change even if there were no fluctuations in the MR measurement. As the effect sizes drop, the p-values tend to increase, as they depend on the combination of both variance and effect size.
	
\end{itemize}

\subsubsection{Additional verification}\label{sec:davide}

\begin{figure}[!htbp]
	\begin{subfigure}{6cm}
		\centering\includegraphics[width=5.5cm]{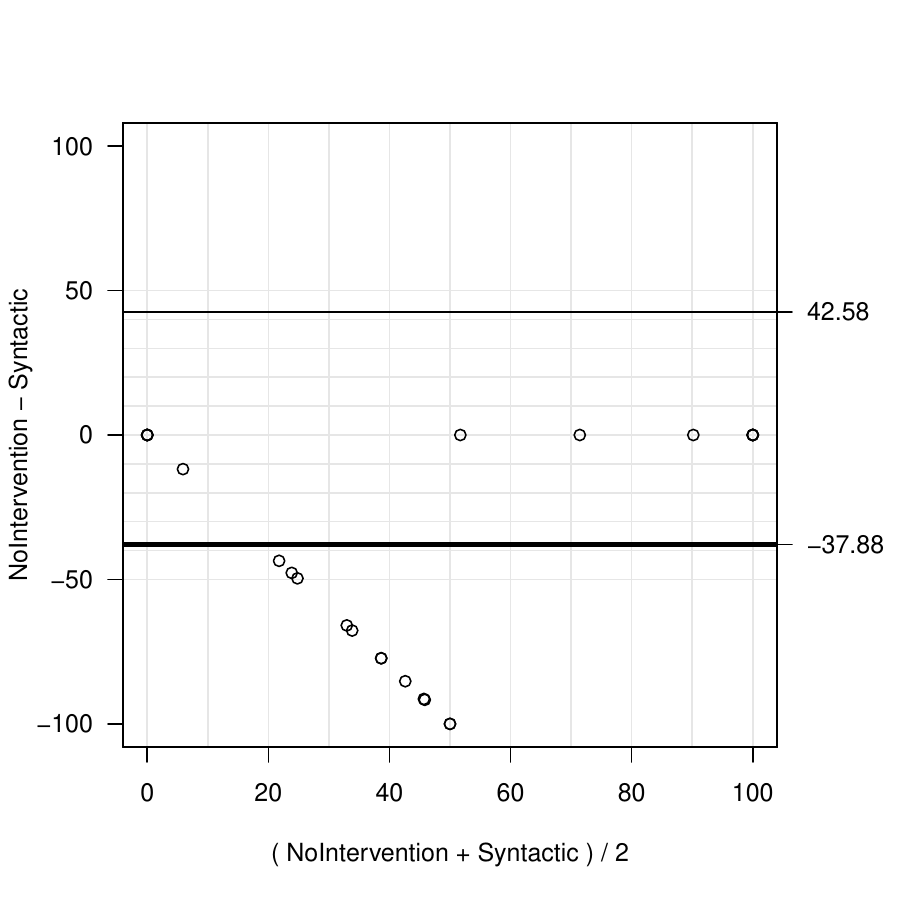}
		\caption{MR}
	\end{subfigure}
	\begin{subfigure}{6cm}
		\centering\includegraphics[width=5.5cm]{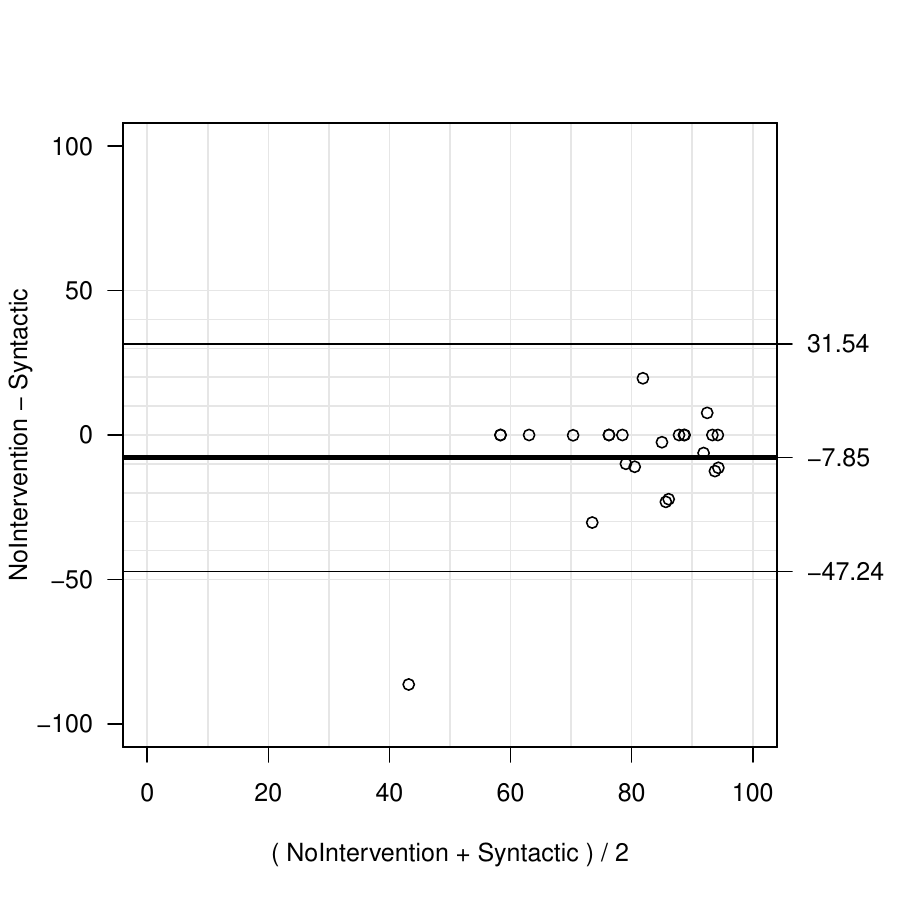}
		\caption{BSK}
	\end{subfigure}
	
	\begin{subfigure}{6cm}
		\centering\includegraphics[width=5.5cm]{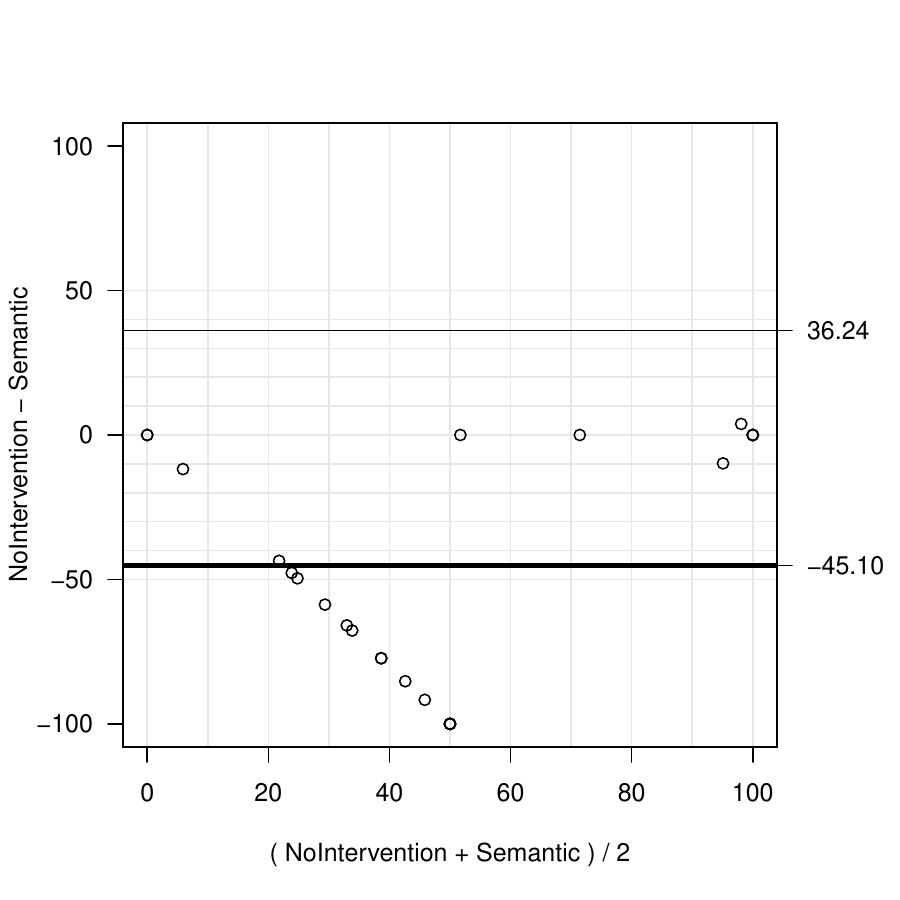}
		\caption{MR}
	\end{subfigure}
	\begin{subfigure}{6cm}
		\centering\includegraphics[width=5.5cm]{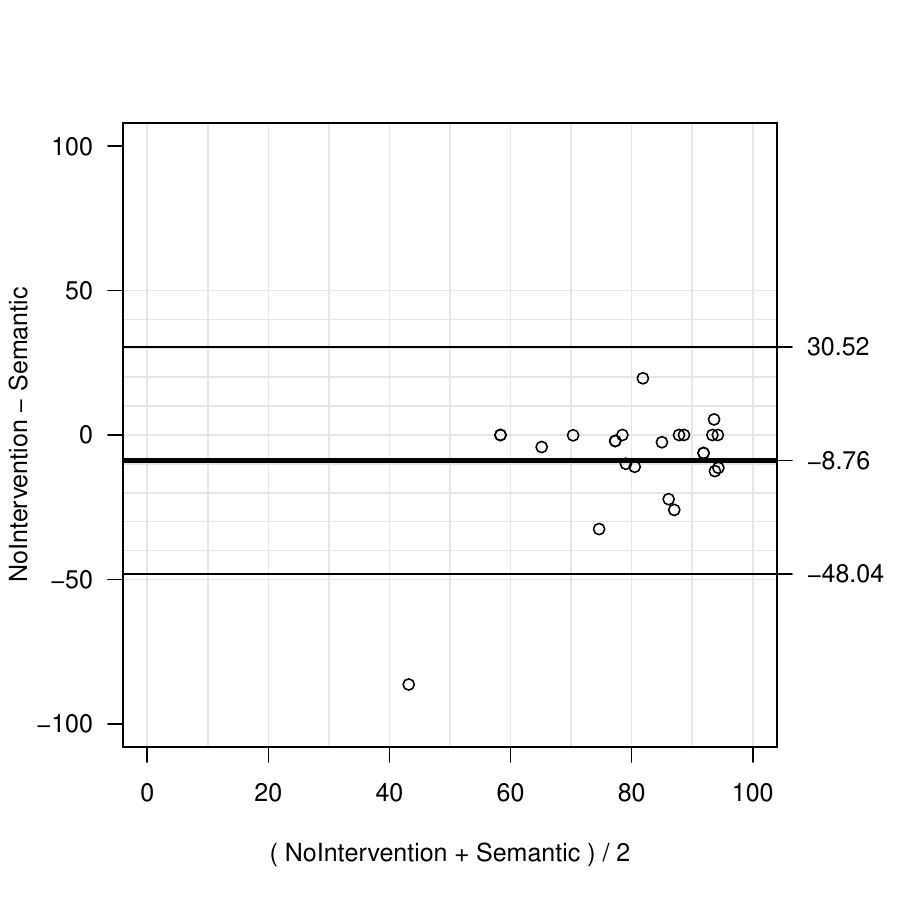}
		\caption{BSK}
	\end{subfigure}
	
	\begin{subfigure}{6cm}
		\centering\includegraphics[width=5.5cm]{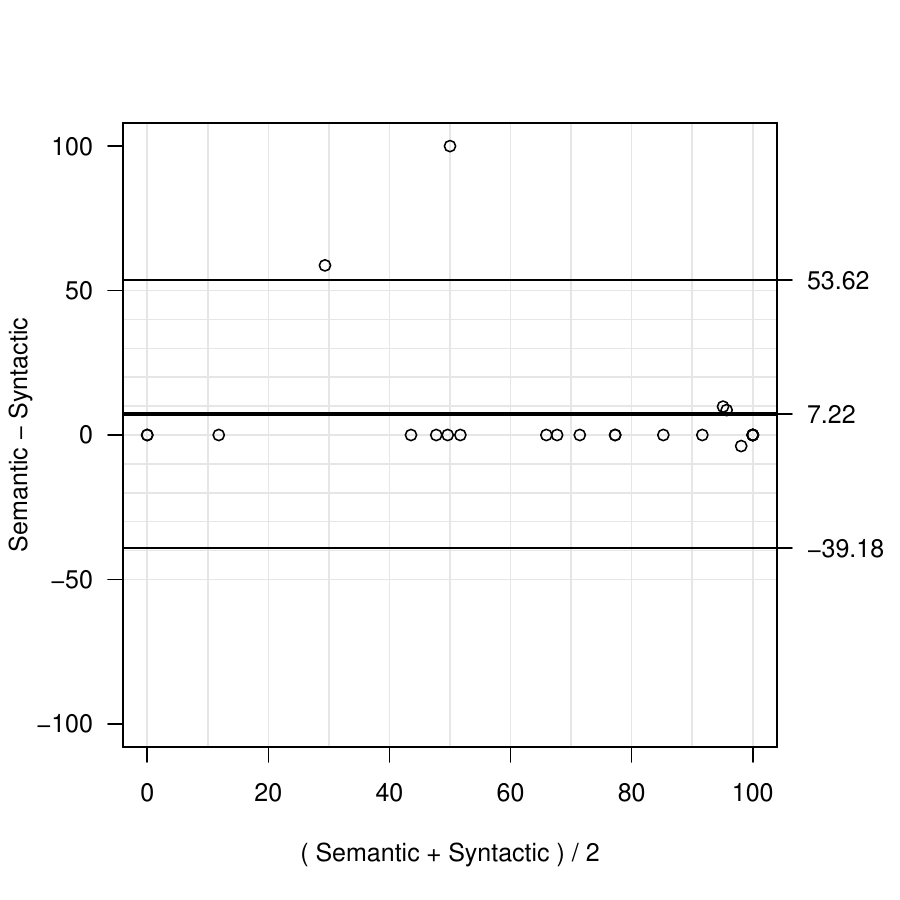}
		\caption{MR}
	\end{subfigure}	
	\begin{subfigure}{6cm}
		\centering\includegraphics[width=5.5cm]{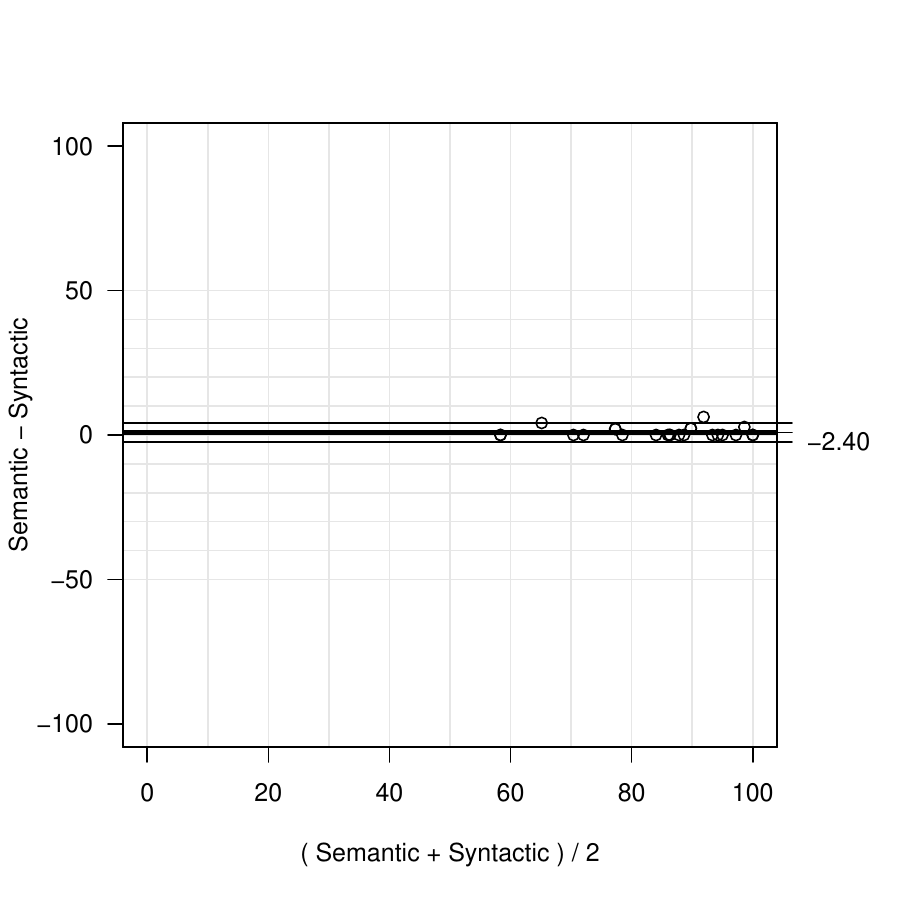}
		\caption{BSK}
	\end{subfigure}

	\caption{Bland-Altman plots for differences between intervention types (measurer: D. Fucci).}
	\label{fig:bland-altman-plots-davide-intervention}

\end{figure}

It might be argued that F. Uyaguari’s measurements merely constitute anecdotal evidence. On this ground, D. Fucci, who was the original measurer of the FS experiment, performed the same exercise as F. Uyaguari. Table \ref{tab:davide-intervention} and Figure \ref{fig:bland-altman-plots-davide-intervention} replicate the analyses reported for the data obtained by F. Uyaguari shown in Table \ref{tab:fernando-intervention} and Figure \ref{fig:bland-altman-plots-fernando-intervention}. It is immediately clear that the results of the statistical analyses and Bland-Altman plots obtained with the measurements made by D. Fucci have the same characteristics as specified above for the measurements made by F. Uyaguari.

\begin{table}[h]
	\small
	\centering
	\caption{Analysis of the QLTY response variable for the three intervention types (measurer: D. Fucci)}
	\label{tab:davide-intervention}
	\begin{tabular}{l c c c}
		\hline
		& No Intervention & Syntactic & Semantic \\
		\hline
		(Intercept)             & $25.55 \; (6.69)^{***}$ & $63.44 \; (5.59)^{***}$ & $70.66 \; (4.99)^{***}$ \\
		TreatmentTDD-greenfield & $50.56 \; (9.46)^{***}$ & $20.53 \; (7.91)^{*}$   & $14.22 \; (7.06)^{*}$   \\
		\hline
		R$^2$                   & $.38$                   & $.13$                   & $.08$                   \\
		Adj. R$^2$              & $.37$                   & $.11$                   & $.06$                   \\
		Num. obs.               & $48$                    & $48$                    & $48$                    \\
		\hline
		\multicolumn{4}{l}{\scriptsize{$^{***}p<0.001$; $^{**}p<0.01$; $^{*}p<0.05$}}
\end{tabular}\end{table}

\subsection{Measurers}\label{sec:measurers}

\subsubsection{Description of measurers}

The third factor that may influence the measurement process is the measurer. In fact, there are numerous references in the literature to the variability introduced by the measurer \cite{Gibbons2002inter, Spring2001intra} to the point that several methods have been developed to calculate the agreement between measurers, published, for example, by Magnusson \cite{magnusson2003handbook, magnusson2007nordest,naykki2012software}, Farrance \cite{farrance2018uncertainty} and Padoan \cite{padoan2017approach}. In the case of the FS experiment, the measurers were, as already mentioned, F. Uyaguari and D. Fucci.

\subsubsection{Impact on statistical analyses} \label{sec:impact_stat_analyses}

The apparent similarity between the measurements by F. Uyaguari and D. Fucci shown in Tables~\ref{tab:fernando-intervention} and \ref{tab:davide-intervention}) is confirmed by the statistical analysis of the FS experiment reported in Table~\ref{tab:analysis-measurers}: 
\begin{itemize}
	
	\item The \textit{TDD-greenfield} effect is very similar for both measurers. For F. Uyaguari, the effect is $29.20$ irrespective of the intervention type applied. For D. Fucci, the effect is $28.44$. The difference between the two effects is a negligible $0.72$ points.
	
	\item The statistical significance is the same in both cases.

\end{itemize}

\begin{table}[h]
\small
\centering
\caption{Analysis of the QLTY response variable for the two measurers (F. Uyaguari and D. Fucci).}
\label{tab:analysis-measurers}
\begin{tabular}{l c c}
	\hline
	& F. Uyaguari & D. Fucci \\
	\hline
	(Intercept)             & $54.88 \; (3.75)^{***}$ & $53.22 \; (3.71)^{***}$ \\
	TreatmentTDD-greenfield & $29.20 \; (5.31)^{***}$ & $28.44 \; (5.25)^{***}$ \\
	\hline
	R$^2$                   & $.18$                   & $.17$                   \\
	Adj. R$^2$              & $.17$                   & $.17$                   \\
	Num. obs.               & $144$                   & $144$                   \\
	\hline
	\multicolumn{3}{l}{\scriptsize{$^{***}p<0.001$; $^{**}p<0.01$; $^{*}p<0.05$}}
\end{tabular}
\end{table}

\subsubsection{Source of the differences} \label{sec:source_differences}

Differences between measurers do not change the value or the sign of the effect, and the statistical significance is unchanged. This suggests that the measurements made on the same code by different measurers are very similar to each other. In the case of TDD experiments that use test cases for quality measurement, the measurer does not appear to be a significant factor of variability. We already mentioned the similarities between the measurements made by F. Uyaguari and D. Fucci in Section \ref{sec:impact_stat_analyses}. The statistical analyses shown in Tables \ref{tab:fernando-intervention} and \ref{tab:davide-intervention} are more or less the same, as are the Bland-Altman plots in Figures \ref{fig:bland-altman-plots-fernando-intervention} and \ref{fig:bland-altman-plots-davide-intervention}.

Figure \ref{fig:bland-altman-plot-davide-fernando} shows a more formal similarity test of the measurements made by the different measurers. Figure \ref{fig:bland-altman-plot-davide-fernando} groups all the intervention types (we discuss what happens when we divide by intervention type later). The variation bounds are approximately +/- 40 points. This value is not surprising, as the data used in Sections \ref{sec:interventions} and \ref{sec:source_differences} \textit{ought} to be the same, and the bounds have the same sources (occasional large differences and/or outliers).

\subsubsection{Impact of the differences on statistical analyses}

\begin{figure}[h]
	\begin{subfigure}{6cm}
		\centering\includegraphics[width=6cm]{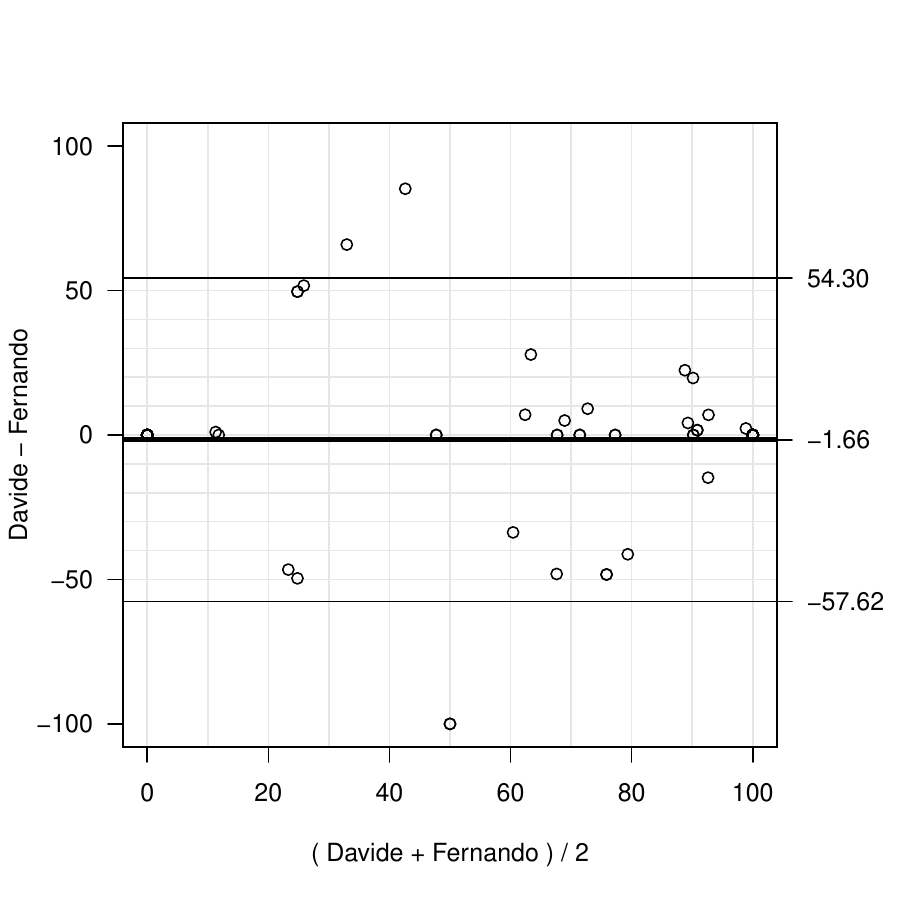}
		\caption{MR}
	\end{subfigure}
	\begin{subfigure}{6cm}
		\centering\includegraphics[width=6cm]{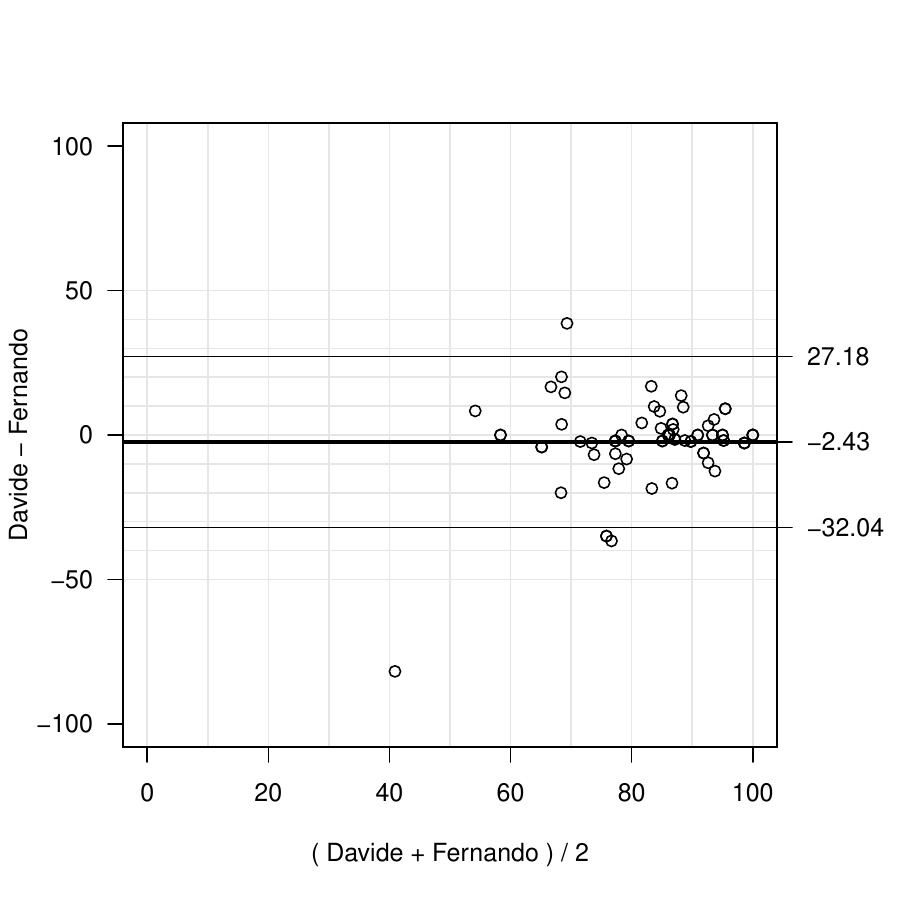}
		\caption{BSK}
	\end{subfigure}

	\caption{Bland-Altman plots for differences between measurers}
	\label{fig:bland-altman-plot-davide-fernando}
\end{figure}

The main result, however, refers to measurer bias. Figure \ref{fig:bland-altman-plot-davide-fernando} shows that bias is practically non-existent (only -2 points for both measurers) and, additionally, the differences are distributed symmetrically (probably randomly) around the central line. In other words, the measurement carried out by one or other measurer does not influence the results of the TDD experiments that use test cases for measuring quality.

\begin{figure}[h]
	\begin{subfigure}{6cm}
		\centering\includegraphics[width=6cm]{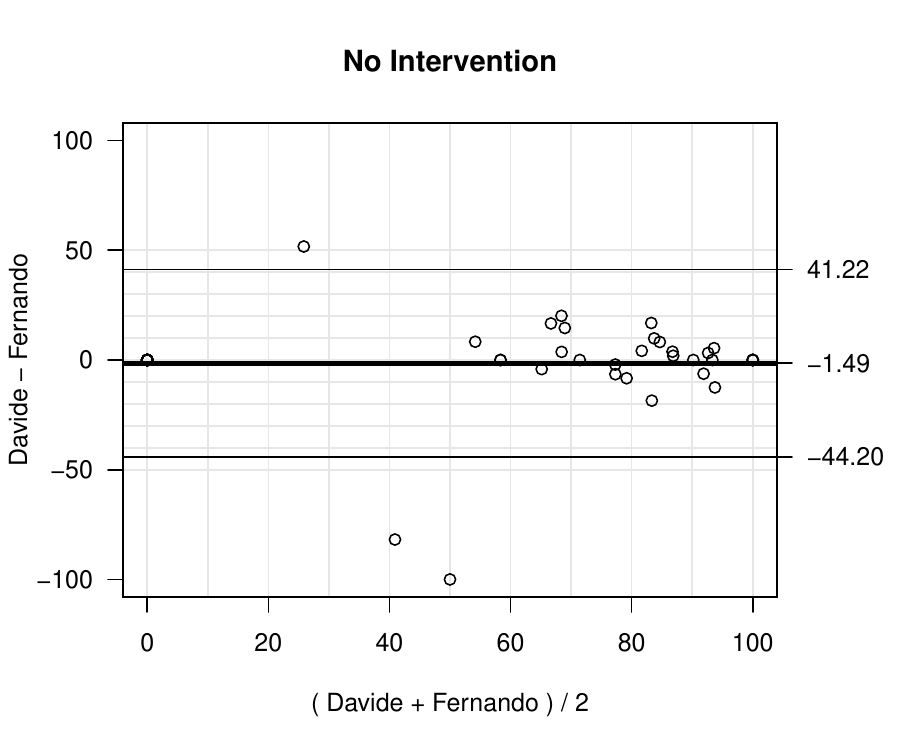}
		\caption{No intervention}
	\end{subfigure}
	\begin{subfigure}{6cm}
		\centering\includegraphics[width=6cm]{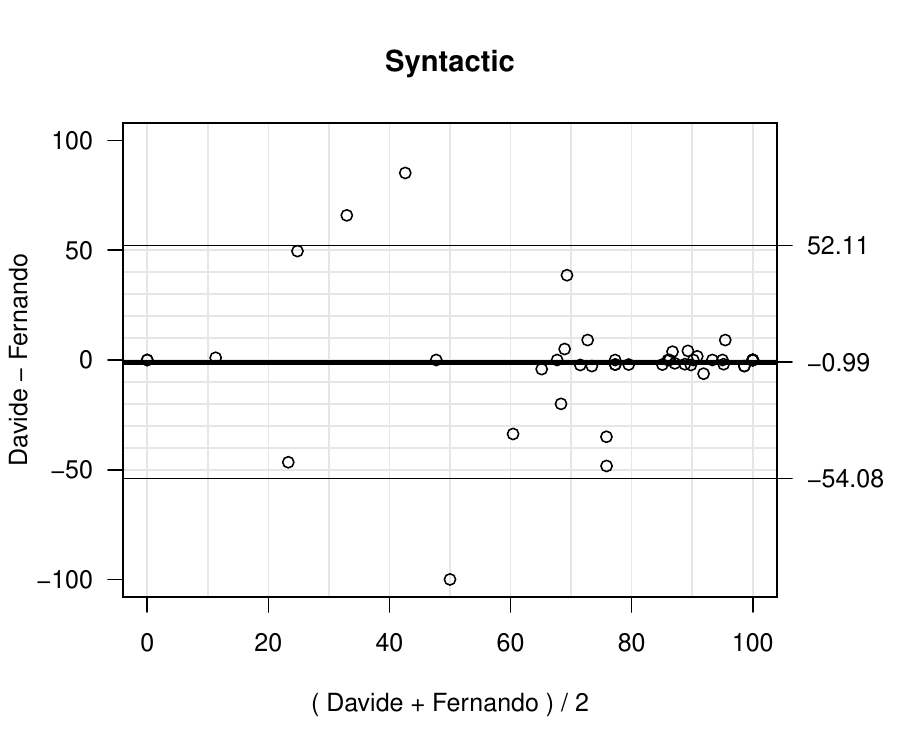}
		\caption{Syntactic}
	\end{subfigure}
	\begin{subfigure}{6cm}
		\centering\includegraphics[width=6cm]{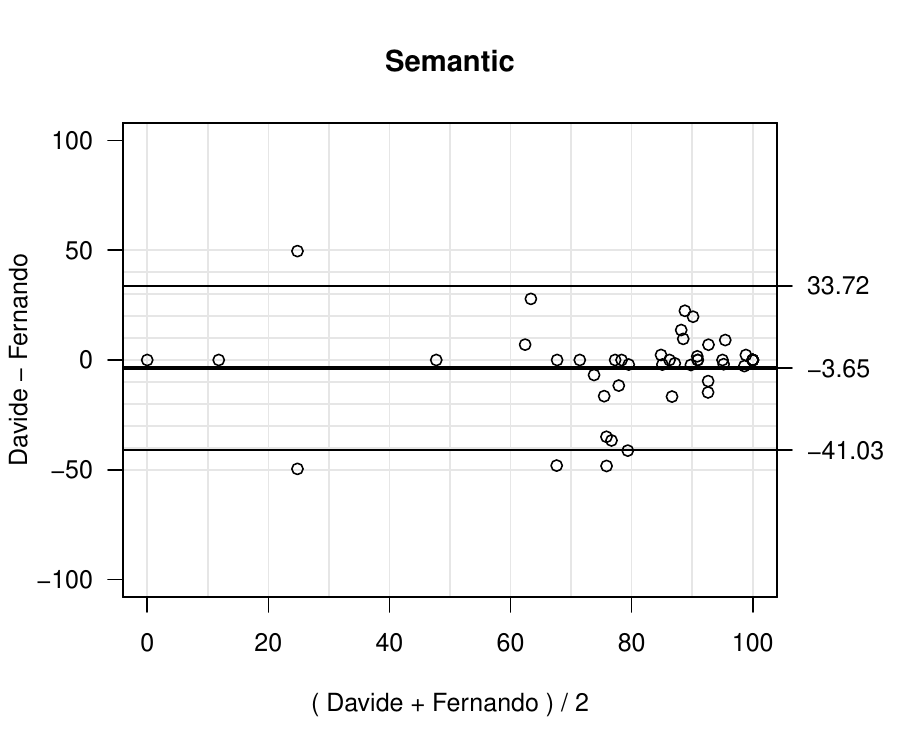}
		\caption{Semantic}
	\end{subfigure}
	
	\caption{Bland-Altman plots for differences between measurers, broken down by intervention type}
	\label{fig:bland-altman-plots-davide-fernando-intervention}
\end{figure}

Finally, we should briefly discuss the differences between measurements broken down by intervention type, which are shown in Figure \ref{fig:bland-altman-plots-davide-fernando-intervention}. As the differences between MR and BSK are symmetric with respect to the central line, and the bias is very low, we show the Bland-Altman plots for both tasks together.

Generally, the Bland-Altman plots in Figure \ref{fig:bland-altman-plots-davide-fernando-intervention} and Figure \ref{fig:bland-altman-plot-davide-fernando} are very similar. However, there are two aspects that should be highlighted:

\begin{itemize}
	
	\item There are differences between measurers when there is \textit{no intervention}. This is noteworthy because a measurer using this intervention type does not manipulate either the subject code or the test cases.
	
	\item The less rigorous intervention types do not produce bigger differences than the \textit{no intervention} type. In fact, \textit{semantic intervention} is the intervention type that causes least differences. There is a difference of only +/- 10 points in the bounds of \textit{no intervention} and \textit{syntactic intervention}.
	
\end{itemize}

Measurers using \textit{no intervention} measurement do not touch the source code. The differences may be caused by human errors by the measurers when enacting the source code loading procedure, running the test cases and transcribing the measurement results.

\section{Systematic Mapping Study}

Our aim is to identify the characteristics of the operationalization components (case studies, experimental task, intervention type and measurer) of the TDD experiments that study external quality and examine how they are reported in order to analyse whether the experimental reports describe the components used to measure the response variable.

\subsection{Procedure}

The research method used to obtain information about the operationalization of TDD and answer the review research questions described below was a SMS conducted according to Kitchenham and Charters’ guidelines~\cite{keele2007guidelines}.

\subsubsection{Review questions}

We stated the following review questions:

\begin{itemize}
	
	\item Q1. How many TDD experiments study external quality using acceptance tests passed?	
	\item Q2. Is the information with respect to the test cases used in the operationalization of TDD experiments that study external quality available?
	\item Q3. Is the information with respect to the experimental task used in the operationalization of TDD experiments that study external quality available?
	\item Q4. Is the information with respect to the code intervention type used in measurement of TDD experiments that study external quality available?	
	\item Q5. Is the information with respect to the measurer of TDD experiments that study external quality available?
\end{itemize}

The procedure enacted to answer the review questions is described below.

\subsubsection{Search process}

For the purpose of answering the review questions, we searched for relevant TDD experiments that study quality published no later than 2021. The search string included the different key words used in the scientific literature. The terms were defined from the following viewpoints:

\begin{itemize}
	
	\item Technique: Test Driven Development, TDD, Test-Driven Development, Test First Development	
	\item Importance factor: Quality
	\item Study types: Experiment
	
\end{itemize}

For search string construction, the logical notation was defined as follows: (T1 or T2, …, or Tn) AND (F1 or F2, …, or Fn) AND (E1 or E2, …, or En). The search string was: \textit{(``Test Driven Development'' OR ``TDD'' OR ``Test-Driven Development'' OR ``Test First Development'') AND ``quality'' AND ``experiment''}.

We used several scientific publisher search engines. The search string was applied to find relevant publications in the online databases used in SE and computer science. The digital databases include IEEE Xplore, ACM Digital Library, SpringerLink, ScienceDirect, most of which were recommended by Brererton et al.~\cite{brereton2007lessons} for conducting systematic literature reviews in SE.

\subsubsection{Study selection}

The study selection was conducted applying a set of inclusion and exclusion criteria. The inclusion criteria are as follows:

\begin{itemize}
	\item The study must address the TDD development technique,
	\item The study reports the experimental results,
	\item The experiment used test cases to measure the external quality response variable. 
\end{itemize}

We excluded studies published in a language other than English.




The preliminary database search returned a total of 1922 publications, including conference proceedings papers and journal articles. We then applied the inclusion and exclusion criteria, which output 20 prospective primary studies. Of the 20 experiments, \textbf{18 use acceptance test cases passed to calculate external quality}. Two quality experiments were not taken into account on the following grounds: the study by Canfora et al. \cite{Canfora2006} measures quality based on the number of implemented test cases, whereas in the paper by Huang and Holcombe \cite{Huang2009}, the external client answers a questionnaire to evaluate quality after using the software for a month.

\subsubsection{Data extraction and analysis process}

From the selected experiments, we extracted the following data:

\begin{itemize}
	\item Study title
	\item Year of publication
	\item Authors
	\item Study type
	\item Context (Industry/Academia)
	\item Control treatment for TDD comparison
	\item Journal/conference where the study was published
	\item Experiment result with respect to external quality
	\item Test cases used
	\item Author of the test cases
	\item Test case generation technique
	\item Name of the experimental task
	\item Description of the experimental task
\end{itemize}

Finally, we studied the data extracted to answer the review questions. The results are reported in Section \ref{sec:resultados}.

\subsection{Results}\label{sec:resultados}

\begin{figure}[h]

	\begin{subfigure}{6cm}
		\centering\includegraphics[width=6cm]{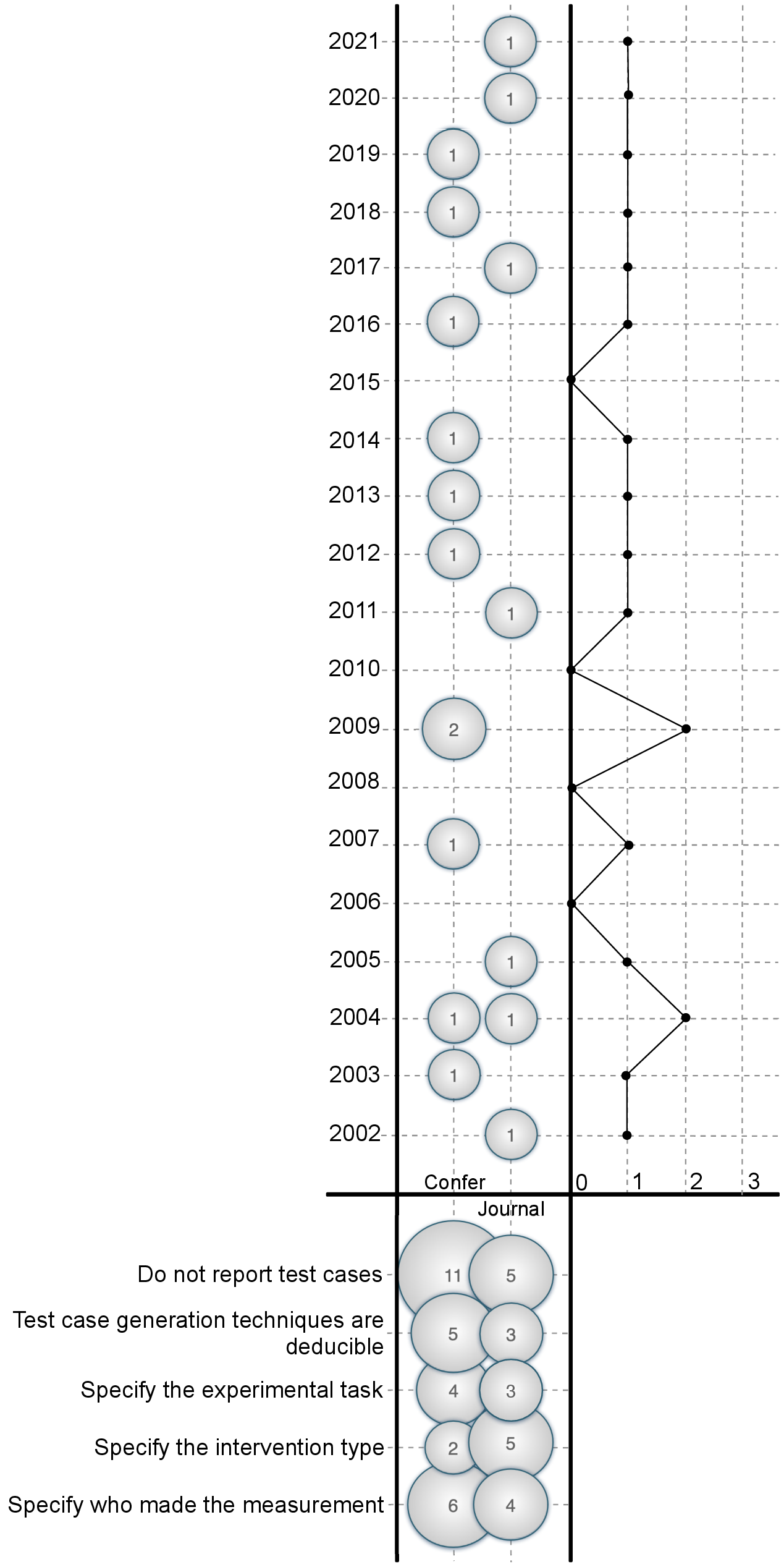}
	\end{subfigure}
	
	\caption{Mapping with the distribution of the primary studies}
\label{fig:accuracy}
\end{figure}

Figure \ref{fig:accuracy} gives an overview of the identified primary studies. The results are segmented into two separate areas in Figure \ref{fig:accuracy}. The first area (left-hand side) basically consists of two XY bubble plots (top and bottom), where the bubbles are placed at the intersections of the following categories: publication type-publication year (top left-hand side) and publication type-information supplied by the experiment (bottom left-hand side). The publication types are conferences and journals. The size of each bubble is determined by the number of primary studies that were classified as members of each category. As the top left-hand side of Figure \ref{fig:accuracy} shows, about two-thirds of the primary studies were published in conferences, which was the preferred medium used by researchers to disseminate their work as of 2012. With respect to the information supplied by the experiment (bottom left-hand side of Figure \ref{fig:accuracy}), just under a third of the experiments specify the experimental task and the intervention type. Additionally, none of the 16 TDD experiments that use acceptance test cases passed to measure external quality report the test cases. Although it is possible to figure out which test case generation techniques are used in just under half of the experiments, they are not exactly specified, defined or deducible in the remainder. On the other hand, just over half of the experiments give an account of who made the measurements. The second area (right-hand side) of Figure \ref{fig:accuracy} shows the number of primary studies by year of publication. Looking at this part of Figure \ref{fig:accuracy}, half of the experiments on TDD that study external quality were reported in the last 10 years.

\subsubsection{TDD experiments that study external quality using test cases}

Table \ref{t:experimentosTDD} shows a list of 18 primary studies. For each of the experiments, it specifies: (i) the primary study identifier (ID); (ii) experimental context (A for academic and I for industrial experiments, respectively), (iii) journal/conference where the primary study was published, (iv) the control treatment for comparison against TDD, (v) the experimental result with respect to quality ((+) the author states that the TDD treatment increases external quality, (-) TDD reduces external quality, (/) there is no difference between the quality results caused by the treatments), (vi) observations, and (vii) the respective reference.

As observed in Table \ref{t:experimentosTDD}, 11 experiments were run in academia and seven, in industry. Of all the experiments, TDD increases external quality in seven (38.8\%), decreases external quality in two (11.11\%), and there is no difference between TDD and the control treatment in nine (50.0 \%). The results are inconclusive and even contradictory.

\begin{table*}[htbp]
	\large	
	\centering
	\caption{TDD experiments that study external quality}
	\label{t:experimentosTDD}
	\resizebox{\textwidth}{!}{%

		\begin{tabular}{p{1.2cm}p{0.7cm}p{3.5cm}p{1.5cm}p{8.0cm}p{3.5cm}p{1.0cm}}
			\hline
			ID & Indu-stry / Academia & Journal/Conference & Control & Experiment result according to author & Observations & Ref.\\ \hline
			[PS11] & A & EASE 2012 & TL & (+) Quality is better with TF than TL. & It is not statistically significant. & \cite{Causevic2012}  \\ \hline
			[PS19] & A & SIGSE 2009 & TL & (/) There is no significant difference between subjects applying TL vs. TDD. & & \cite{Desai2009} \\ \hline
			[PS25]  & A & IEEE Transactions on Software Engineering & ITL & (/) There is no significant difference between ITL \& TDD. ITL: 85\% quality, TDD: 83\% quality, p-value: 0.25 & & \cite{Erdogmus2005}\\ \hline
			[PS30] & A  & ESEM 2013 & ITL   & (/) Small effect size: 0.11, p-value: 0.53  & The null hypothesis is not rejected. & \cite{Fucci2013} \\ \hline			
			[PS108] & A  & ESEM 2016 & ITL   & (/) Very small difference between TDD \& ITL wrt quality & The difference is not statistically significant. & \cite{Fucci2016} \\ \hline		
			[PS37]  & I & Information and Software Technology & Waterfall & (+) TDD increases quality by 18\% over waterfall. &  & \cite{George2004}\\ \hline
			[PS38] & I & METRICS 2004 & TL & (+) TDD increases quality wrt TL. &  & \cite{Geras2004} \\ \hline
			[PS40] & A & ESEM 2007 & Waterfall & (/) Quality increased due to testing effort not to development technique.  &  & \cite{Gupta2007} \\ \hline
			[PS78] & A & IEE Proceedings - Software & Waterfall & (/) TDD does not increase program reliability.  & Reliability is measured with acceptance cases passed. & \cite{Muller2002} \\ \hline
			[PS80] & I & EASE 2014, ACM & ITL   & (/) Similar number of acceptance tests passed.  & The difference is not statistically significant. & \cite{Munir2014e} \\ \hline
			[PS82] & A & Eurocon 2003 & ITL & (-) Small improvements for ITL. TDD: mean value of 92.6/120 acceptance tests passed; ITL: 95.1/120 & There is no statistically significant difference. & \cite{Pancur2003} \\ \hline
			[PS83]  & A & Information and Software Technology & Micro ITL & (+) The TDD effect is small and positive wrt the percentage of acceptance test cases. & There is no statistically significant difference. & \cite{Pancur2011} \\ \hline
			[PS98]  & A & ITNG & ITL & (+) ITL has 39\% more defects than TDD (number of recorded defects). & & \cite{Vu2009} \\ \hline
			[PS115]  & I & ICCIKE 2019 & BDD & (+) TDD increases external quality. & & \cite{dookhun2019assessing} \\ \hline
			[PS116]  & A & IEEE Transactions on Software Engineering & TL & (+) TDD has an edge over TLD in terms of code quality (1.8 fewer bugs). & It is statistically significant. & \cite{papis2020experimental} \\ \hline
			[PS117]  & I & APSEC 2018 & YW \& ITL & (/) TDD slightly outperforms YW \& ITL. & It is not statistically significant. & \cite{santos2018improving} \\ \hline
			[PS118]  & I & Empirical Software Engineering & ITL & (/)  There is no statistically significant difference between the quality of the work performed by subjects for both treatments. &  & \cite{Tosun2017} \\ \hline
			[PS119]  & I & IEEE Transactions on Software Engineering & ITL & (-) The subjects produce better quality code with ITL than with TDD where the subtasks are divided into user stories, coded and tested. &  & \cite{tosun2019investigating} \\ \hline
			
		\end{tabular}	
	}
	\raggedright
	\small{TF=Test First; TL = Test Last; ITL = Iterative Test Last, BDD = Behaviour-driven development; YW = Your way}
	
\end{table*}

\normalsize

This study focuses on experiments where quality was measured  using acceptance test cases passed. The measurement procedure of these experiments is shown in Figure \ref{fig:figura-procesos-medicion}.

\subsubsection{Test cases}

Table \ref{t:CasosDePrueba} shows the information extracted from the primary studies on test case generation. For each primary study, it specifies: (i) its identifier, (ii) the experimental context, (iii) whether the test cases are reported, (iv) who ran the test cases, (v) the technique used to generate the test cases, and (vi) the respective reference.

\begin{table*}[h]
	\large
	\centering
	\caption{Test cases in TDD experiments}
	\label{t:CasosDePrueba}
	\resizebox{\textwidth}{!}{%

		\begin{tabular}{p{1.2cm}p{1.0cm}p{2.2cm}p{6.0cm}p{3.7cm}p{1.2cm}}
			\hline
			ID & Indu-stry/
				Academia
			 & Are the test cases reported?     & Who generated the test cases? & Test case generation technique used & Ref.  \\ \hline 
			
			[PS11] & A & NO    & We deduced that it was the subjects. & We deduced that technique is ad hoc. & \cite{Causevic2012} \\  \hline			
			[PS19] & A & NO    & The instructor created the test case suite. & We deduced that technique is ad hoc. & \cite{Desai2009} \\  \hline
			[PS25] & A & NO    & Black box acceptance tests were written by the researcher & Unspecified. & \cite{Erdogmus2005} \\ \hline
			[PS30] & A  & NO    & They were created by the researcher. & We deduced that technique is ad hoc. & \cite{Fucci2013} \\ \hline
			[PS108] & A & NO    & We deduced that they were created by the researcher. & We deduced that technique is ad hoc. & \cite{Fucci2016} \\ \hline			
			[PS37] & I & YES, in the thesis \cite{George2002} but NOT in the primary study & They were created by the researcher. & We deduced that technique is ad hoc. & \cite{George2004} \\ \hline
			[PS38] & I & NO    & There are developer test cases created by the developer and client test cases created by the researcher. & Unspecified. & \cite{Geras2004} \\ \hline			
			[PS40] & A & NO    & They were created by the researcher. & Unspecified. & \cite{Gupta2007} \\ \hline			
			[PS78] & A & NO    & Unspecified & We deduced that technique is ad hoc. & \cite{Muller2002} \\ \hline
			[PS80] & I & NO    & Unspecified & Unspecified. & \cite{Munir2014e} \\ \hline
			[PS82] & A & NO    & Unspecified & We deduced that technique is ad hoc. & \cite{Pancur2003} \\ \hline
			[PS83] & A & NO    & They were created by the researcher. & We deduced that technique is ad hoc. & \cite{Pancur2011} \\ \hline
			[PS98] & A & NO    & Unspecified & Unspecified. & \cite{Vu2009} \\ \hline
			[PS115] & I & NO    & Unspecified & Unspecified. & \cite{dookhun2019assessing} \\ \hline
			[PS116] & A & YES, in the replication package & We deduced that they were created by the authors. & Unspecified. & \cite{papis2020experimental} \\ \hline
			[PS117] & I & NO & Unspecified & Unspecified. & \cite{santos2018improving} \\ \hline			
			[PS118] & I & NO & Unspecified & Unspecified. & \cite{Tosun2017} \\ \hline
			[PS119] & I & YES & They were created by the researchers. & Unspecified. & \cite{tosun2019investigating} \\ \hline						
		\end{tabular}	
	}
	
\end{table*}

\normalsize

As Table \ref{t:CasosDePrueba} shows, test cases are reported in three experiments and not detailed in 15, that is, 16.66 \% and 83.33 \%, respectively. The authors fail to specify who generated the test cases in seven experiments (38.88 \%). No information about the technique used to generate the test cases is supplied in 10 experiments (55.55 \%). None of the 18 experiments explicitly report the name of the technique used to generate the test cases that are used as an instrument for measuring quality.

Based on the analysis of the primary studies, we find that there are differences between the case studies and, generally, between the instruments that the researchers use in the experiments to measure the same constructs . The percentage differences appear to be small, but, in actual fact, we deduced most of the information shown in Table \ref{t:CasosDePrueba}. The authors provide little information on the characteristics of the test cases used in the measurements. Additionally, they offer little or no information on how the test cases were generated, where they were obtained or who created them. None of the analysed experiments explicitly explain whether they use a specific test case generation technique. Eight of the analysed experiments (44.44 \%) use AH, which is essentially tantamount to the researcher’s best judgement, instead of a formal test case generation technique.

\subsubsection{Experimental task}

The experimental task is another focus of this study. In the 18 experiments that study quality, we identified the name of the task used and determined whether or not the author reports the task specification.

As Figure \ref{fig:figura-procesos-medicion} shows, the experimental task is another important operationalization component in TDD. Table \ref{t:Tarea} reports the characteristics of the experimental task used in each of the 18 primary studies. For each of these experiments, it specifies: (i) their identifier (ID), (ii) the experimental context, (iii) whether or not the BSK experimental task is used, (iv) the name of the experimental task, (v) whether or not it reports the experimental task specification, and (vi) the respective reference.

According to Table \ref{t:Tarea}, three experiments (16.66\%) fail to specify the name of the experimental task, and 11 experiments (61.11\%) do not report the experimental task specification. Bowling Scorekeeper (BSK)\footnote{Bowling Scorekeeper by C. Martin~\cite{BSK}} was the experimental task used in 10 experiments (55.55\%). Five experiments used a task other than BSK (27.77\%). Only three primary studies (16.66\%) explicitly report the experimental task specification used in the experiment, and three primary studies (16.66\%) reference other documents that are available in a repository containing the specification.

\begin{table*}[htbp]
	\normalsize
	\centering
	\caption{Experimental task in the TDD experiments that study external quality}
	\label{t:Tarea}
	\resizebox{\textwidth}{!}{%
		
		\begin{tabular}{p{1.0cm}p{0.8cm}p{1.8cm}p{6.0cm}p{2.0cm}p{1.0cm}}
			\hline
			ID & Indu-stry/
				Academia
			 & Does it use BSK?     & Name of experimental task & Does it report the specification of the experimental task? & Ref.      \\ \hline 
			
			[PS11] & A & YES & BSK & NO & \cite{Causevic2012}    \\ \hline			
			[PS19] & A & Unspecified & It mentions, but does not specify, 7 projects. & NO & \cite{Desai2009} \\ \hline 
			[PS25] & A & YES    & BSK & NO & \cite{Erdogmus2005} \\ \hline 
			[PS30] & A & YES    & BSK & NO & \cite{Fucci2013} \\ \hline 
			[PS108] & A & YES   & BSK & NO & \cite{Fucci2016} \\ \hline 						
			[PS37] & I & YES    & Bowling Game (similar to BSK) & YES, in the thesis  & \cite{George2004} \\ \hline 
			[PS38] & I & NO    & Program A – Registering a new project
			Program B – Recording time against a project
			& YES & \cite{Geras2004} \\ \hline 			
			[PS40] & A & NO    & Student registration system / Simple ATM system & YES &  \cite{Gupta2007} \\ \hline	
			[PS78] & A & NO    & GraphBase, related to graphs & NO &  \cite{Muller2002}  \\ \hline 
			[PS80] & I & YES    & BSK & YES, on the experiment web page & \cite{Munir2014e}  \\ \hline 
			[PS82] & A & Unspecified & Unspecified & NO & \cite{Pancur2003}   \\ \hline 
			[PS83] & A & NO    & Distributed database server with built-in data replication mechanism, implementation of a chat server & NO & \cite{Pancur2011}   \\ \hline 
			[PS98] & A & Unspecified & Unspecified & NO & \cite{Vu2009} \\ \hline 
			[PS115] & I & YES & String calculator, BSK & NO & \cite{dookhun2019assessing} \\ \hline 	
			[PS116] & A & NO & 1) Propose a data structure to hold information about daily gas flow for every hour in local time. 2) Assuming all the required data for the structure designed in Task 1 is stored in a text file, propose a method of verifying its correctness. The participants assume the availability of a method GetNumberOfHoursInDay, which returns an integer and accepts a single DateTime parameter. & Partially & \cite{papis2020experimental} \\ \hline  
			[PS117] & I & YES & BSK, MR, Spreadsheet & YES, in a link & \cite{santos2018improving} \\ \hline 	
			[PS118] & I & YES & MR API, BSK, MusicPhone & NO, outlines tasks  & \cite{Tosun2017} \\ \hline
			[PS119] & I & YES & MR API, BSK & YES & \cite{tosun2019investigating} \\ \hline  				
		\end{tabular}	
	}
	
\end{table*}

\normalsize

\subsubsection{Intervention type}

When the measurer applies the test cases to obtain the value of the response variable, the source code of the experimental task developed by the subjects must be connected to the test cases. Source code intervention is often necessary, which is carried out at the measurer’s discretion. We gathered information on the type of intervention carried out by the measurer on the source code during measurement in the 18 experiments that study quality.

The last but one column in Table \ref{t:Medidor} shows the type of source code intervention in each of the 18 primary studies. Code intervention was not reported in 11 experiments (61.11\%), there was no source code intervention in two experiments (11.11\%), the subjects in three experiments (16.66\%) modified the source code after running the test cases in order to correct errors, and the researcher of one experiment (5.55\%) stated that changes were made so that the code executed successfully.

\begin{table*}[h]
	\small
	\centering
	\caption{Measurer and type of code intervention in TDD experiments that study external quality}
	\label{t:Medidor}
	\resizebox{\textwidth}{!}{%
		
		\begin{tabular}{p{1.0cm}p{1.0cm}p{3.5cm}p{4.5cm}p{1.0cm}}
			\hline
			ID & Indu-stry/
				Academia
			 & Who made the measurements?     & Source code intervention during measurement & Ref.      \\ \hline 
			
			[PS11] & A & Unspecified & Unspecified & \cite{Causevic2012}    \\ \hline			
			[PS19] & A & Researchers & Unspecified & \cite{Desai2009} \\ \hline 
			[PS25] & A & Unspecified & Test cases were run automatically at the end of the experiment. We deduced that there was no source code intervention. & \cite{Erdogmus2005} \\ \hline 
			[PS30] & A & Unspecified & Unspecified & \cite{Fucci2013} \\ \hline 
			[PS108] & A & Researcher (FU) & Unspecified & \cite{Fucci2016} \\ \hline 						
			[PS37] & I & Unspecified & Unspecified & \cite{George2004} \\ \hline 
			[PS38] & I & Subjects & The test cases are run automatically. We deduced that there was no source code intervention. & \cite{Geras2004} \\ \hline 			
			[PS40] & A & Subjects & The test cases were run, and the errors were observed and corrected. &  \cite{Gupta2007} \\ \hline	
			[PS78] & A & Subjects & The subjects corrected the defects. &  \cite{Muller2002}  \\ \hline 
			[PS80] & I & Researchers & Unspecified & \cite{Munir2014e}  \\ \hline 
			[PS82] & A & Unspecified & Unspecified & \cite{Pancur2003}   \\ \hline 
			[PS83] & A & Subjects & The subjects modified the code until the test case pass rate was 100\%. & \cite{Pancur2011}   \\ \hline 
			[PS98] & A & Committee composed of representatives of subject groups &  Unspecified & \cite{Vu2009} \\ \hline 
			[PS115] & I & Unspecified & Unspecified & \cite{dookhun2019assessing} \\ \hline 
			[PS116] & A & Measured automatically using web services at the end of each iteration & Unspecified & \cite{papis2020experimental} \\ \hline 			
			[PS117] & I & Unspecified & Unspecified & \cite{santos2018improving} \\ \hline
			[PS118] & I & Unspecified & The measurement description does not state that there is code intervention. & \cite{Tosun2017} \\ \hline 	
			[PS119] & I & Researcher & Changes were made so that the code ran successfully without interfering with the code delivered by the subject. & \cite{tosun2019investigating} \\ \hline											
		\end{tabular}	
	}
	
\end{table*}

\normalsize

\subsubsection{Measurers}

The measurer applies the test cases to obtain the value of the response variable. 

Table \ref{t:Medidor} specifies for each primary study: (i) its identifier (ID), (ii) the experimental context, (iii) the measurer, (iv) data on the source code intervention when making the measurements, and (v) the respective reference.

According to Table \ref{t:Medidor}, we find that eight experiments (44.44\%) do not specify who does the measuring, four experiments (22.22\%) identify the researchers as the measurers, the subjects did the measuring in four experiments (22.22\%), a committee was formed to measure quality in one experiment (5.55\%), and a weekly measurement was made automatically at the end of each iteration in one experiment (5.55\%).

\subsection{Review conclusion}

With respect to the analysed measurement process components, such as test cases used to measure external quality, experimental task, intervention type and measurers, the researchers do not report or only partially report the characteristics of these components. We find that the characteristics of the instruments —experimental task, development environment, test case generation, test case generation technique, measurers, code interventions— used by researchers in each experiment may well differ. Although there are SE experiment reporting guidelines, researchers do not include all the information in experiment reports.

\section{Validity threats}\label{sec:validez}

This section discusses the validity threats with respect first to the quantitative research and then to the qualitative study. With regard to the quantitative research, we take into account the following:

For the purpose of assuring conclusion validity, we took the following precautions:

\begin{itemize}
	
	\item We used several methods of comparison, namely the concepts established in ISO 5725 \cite{ISO5725} and Bland-Altman plots \cite{Bland1986}.
	
	\item The conclusions are based on a total of 288 measurements made for this study.
	
	\item Data capture and analysis were largely automated to prevent problems of transcription and manipulation. R and Sweave were used to make and integrate all the calculations.
	
\end{itemize}

With regard to internal validity, the following actions were carried out to assure that the conclusions capture the effect of the independent variable on the dependent variable:

\begin{itemize}
	
	\item The measurements were taken independently, based on the original source code created by the experimental subjects, that is, code was not reused to rule out undesired dependencies across measurements.
	
	\item The measurers (i.e., F. Uyaguari and D. Fucci) acted independently without communicating with each other.
	
	\item The data used in this study are not the first measurements made by the measurers F. Uyaguari and D. Fucci. These measurers had already made measurements as part of other experiments.

\end{itemize}

To assure that the resulting conclusions are linked to the independent and dependent variable operationalizations and not to underlying constructs, we took precautions with respect to the construct validity:

\begin{itemize}
	
	\item In this study, we studied differences between the measurements of the quality response variable using test cases only.
	
	\item For each measurement method component, that is, the independent variables, we study different variants:
	\begin{itemize}
		\item Two types of test case generation strategies
		\item Two different measurers
		\item Three intervention types.
	\end{itemize}
	
\end{itemize}

To assure that the research results can be generalized to other populations, situations or moments of time, we took the following precautions with respect to external validity:

\begin{itemize}
	
	\item To increase the possibility of generalizing results, we measured experiments carried out by professional subjects that are representative of industry (FS).
	
	\item We used representative test case suites created by different people. The first AH test suite was used to measure experiments in industry within the ESEIL project \cite{juristo2016experiences}. The second suite was generated as part of a master thesis using the EP technique \cite{upm44268}.
	
	\item The measurements were made by representative measurers: two PhD students (at measurement time) with experience in the software area. 
	
\end{itemize}

With respect to the qualitative part, note that this study focused on TDD. The research method used was a SMS. In order to increase the study validity, we identified relevant experiments on TDD within a range of databases like IEEE Xplore, ACM Digital Library, SpringerLink and ScienceDirect used to answer the research questions.

In order to carry out a good quality SMS, the research was conducted by a team of researchers. The search string, primary study selection and data extraction was carried out by the researcher F. Uyaguari and reviewed by another three of the paper authors to assure that the procedure complied with Kitchenham and Charters’ guidelines. The results were synthesized by all four researchers through negotiation.

Finally, there are several frameworks for evaluating the quality of systematic reviews. We used the evaluation criteria proposed by Thompson et al. \cite{Thompson2012} because they are domain independent and easy to use. Table \ref{t:Evaluacion} reports the results of the evaluation of the study, which concludes that the literature review results are reliable.

\begin{table*}[h]
	\scriptsize
	\centering
	\caption{Evaluation criteria used for the systematic review}
	\label{t:Evaluacion}
	\resizebox{\textwidth}{!}{%
		
		\begin{tabular}{p{3.5cm}p{5.5cm}}
			\hline
			Evaluation criteria & Evaluation of review questions \\ \hline 
			
			Was the bibliographic search exhaustive?  & Yes, although Scopus and WoS were missing, the review includes the four major publishers of SE articles (IEEE, ACM, SpringerLink and Elsevier).\\ \hline
			
			Were the criteria used to select articles for inclusion appropriate? & Yes, we used appropriate criteria for searching for articles that report the results of experiments on TDD that study external quality using acceptance tests passed. \\ \hline		
			
			Were the studies included valid enough to answer the stated research questions? & Yes, only controlled experiments were selected. \\ \hline
			
			Were the results similar from one study to another? & Yes, they are articles that are reported in sufficient detail and published in journals/conferences where supplementary information is included in web appendices. \\ \hline

		\end{tabular}	
	}

\end{table*}

\normalsize

\section{Discussion}\label{sec:discusion}

\subsection{Main results of the quantitative analysis}

The results reported in Section 4 highlighted that the measurement process components (test suites, intervention types and measurers) have a powerful influence on the results of both measurements and statistical analyses.

Firstly, let us discuss which measurement method components may threaten the validity of the statistical analyses of the experiments, which was one of the primary aims of this study. With regard to TDD, which is the area explored by the experiments used in the analyses, our observations suggest that:

\begin{itemize}
	
	\item The test suites have a major impact on both the measurements and the statistical analyses of the experiments.
	
	\item The intervention type has an impact on the measurements, and a more limited impact on the statistical analyses of the experiments.
	
	\item The measurers have an impact on the measurements but do not influence the statistical analyses of the experiments.
	
\end{itemize}

The more positive measurements yielded by AH are the source of the differences between the measurements made using different test cases. They have an impact on the influence of the treatment (TDD) to the point that they change the sign of the effect, confirming the results obtained by Dieste et al. \cite{dieste20}. 

With respect to intervention type, the impact on the statistical results is reflected by the fact that, as source code intervention increases, the treatment effect decreases with respect to both the response variable value and statistical significance. The following example illustrates the above statement. We know that syntactic intervention is a more rigorous intervention type than semantic intervention, which means that the QLTY values are lower in the first than in the second case. This is illustrated to perfection in Figures \ref{fig:bland-altman-plots-fernando-intervention} and \ref{fig:bland-altman-plots-davide-intervention}, which compare the two intervention types. This variation modifies the effect on the statistical analysis, because all the codes, irrespective of whether they were developed with ITLD or TDD or simply MR or BSK, are equally affected by low QLTY values. While, unlike the test cases, the sign of the effect is unchanged, it is important to bear this issue in mind to standardize the measurement procedure that is being applied to operationalize the experiments.

The conclusions of the experiment are unchanged irrespective of whether the results of an experiment are measured by one or other measurer or the code delivered by the subjects is unmanipulated or manipulated differently for the purpose of connection to the measurement test suite.

Caution should be exercised when generalizing. However, there are some recommendations that are probably generally applicable:

\begin{itemize}
	
	\item Measurement using test suites is largely automated. It is true that the measurers have to connect the code developed by subjects to the test suites using a series of predefined intervention types. Once this step is completed, however, measurement is made automatically. This process is largely equivalent to measurement in chemistry: the sample is prepared for measurement, and special measurement apparatus is immersed in the solution. The intervention types (equivalent to sample preparation in chemistry) are somewhat but not overly subjective. Therefore, we recommend that, irrespective of the particular branch of science, the measurement method must, whenever possible, be automated.
	
	\item Measurement is automated in TDD experiments by using test suites. However, as reported in Section 5, test suites are the most troublesome measurement method component because they behave differently with respect to AH and EP. The question then is, can these behavioural differences be avoided?
	
	In principle, we believe that they can be avoided in two different ways:

	\begin{itemize}
		
		\item Performing a pilot experiment, which is capable of testing measurement reproducibility \cite{dieste20} and, if necessary, adjusting test suites.
		
		\item Running a reproducibility analysis using reference code. This code could be composed of variants of the MR and BSK tasks that implement some, but not all, user stories.
		
	\end{itemize}
	
	The second option appears to be more preferable on three grounds:
	
	\begin{enumerate}
		\item It requires less effort.
		\item It assures more control over measurements.
		\item The comparison of two measurements is unnecessary, as a reference value (QLTY associated with the reference code) is available.
	\end{enumerate}
	
	In other words, the measurement instrument (test suites) should not be generated based exclusively on the fundaments of theory but must be empirically validated using reproducibility analysis \cite{dieste20}.

	\item One way of improving measurement uniformity is, of course, to take the subjectivity out of the application of the intervention types. As discussed in Section 5 above, the no intervention type is dysfunctional, and, as such, is not an option. We do not know whether this applies in other areas, but it is highly likely. One credible simile are student examinations. For example, have you never hastily marked a student’s exercise 0 and then, on closer reading, found it to be satisfactory or even brilliant? Alternatively, have you never rejected an article during a peer review process, which was finally accepted? A cursory examination does not appear to be the best possible evaluation strategy.
	
	In the field of TDD, one possible way of eliminating measurer bias would be to force subjects to respect the experimental task APIs. Intervention is required primarily due to API modifications. However, it is questionable whether you can oblige a subject to respect an API with which they disagree. It would perhaps be preferable to limit measurer access to the code delivered by the experimental subjects, giving them access to the API alone. This could reduce possible decision-making bias.
	
	Generally, we recommend the use of blinding. The measurers should not have any preference with respect to the code delivered by the subjects (or any other product resulting from the experimental sessions), the applied treatment or, even, the measurement results. This would minimize measurement subjectivity and bias.

	\item It was determined in Section 5 that statistical analysis was not affected when the measurement method components behave consistently with respect to all the factor levels. One much simpler way of testing whether the statistical analysis was affected would be to run the analyses using different measurements and compare the results. However, we do not advise the experimenters to make post hoc decisions. Any decision should precede the statistical analysis of the results (except, of course, the choice of the method (AIC, BIC, etc.) to study model fit).
	
\end{itemize}

Different test case suite and measurer behaviour with respect to the MR and BSK tasks is a source of problems with measurements and measurement analysis. The use of two tasks is only necessary in within-subjects experiments, where the subjects apply more than one treatment to different tasks. In the case of between-subjects experiments, where there is only one experimental task, behaviour is bound to be the same, and test suites and intervention type have no harmful effects.

\subsection{Main results of the qualitative study}

As observed in the scientific literature, TDD experiments that study quality repeatedly yield different, and sometimes even contradictory, results (for example, George and Williams \cite{George2004} claim that TDD favours an 18\% quality increase with respect to the waterfall method, whereas Huang and Holcombe \cite{Huang2009} claim that there is a very small difference, which is not statistically significant). In some secondary studies \cite{Causevic2011, Rafique, Bissi2016}, the researchers look for explanations in the experimental context (industry or academia) and study rigor \cite{Munir2014}. However, these studies fail to take into account other factors, such as components relative to the instrumentation, which could be having an effect on the experimental results.

Researchers only partially report the instrumentation of both the test cases used to measure the external quality and the experimental task. We found that different researchers may very well use instruments —experimental task, test cases used for measurement, test case generation technique, etc.— with different characteristics in their experiments. Even though they have access to guidelines and recommendations for reporting experiments in SE, the authors do not include all the information in the experiment reports. The authors tend to partially report instrumentation-related issues in TDD experiments.

If the researchers do not report the information on the experiments in full, it is impossible to evaluate the external validity of the results or deduce what influence this may have on the experimental results. Different instrument characteristics will probably lead to different experimental results, which are being overlooked by researchers. Access to more detailed information about the test cases and experimental task, code intervention type and other components would be useful for studying the influence of instrumentation on the experimental results.

Additionally, the failure to record these measurement process components is an obstacle to replication, and replication underpins the formation of families of experiments to obtain more reliable results. If we had information on measurement process operationalization, it would help effective experiment replication, assuring that the replication results were not influenced by the factors observed in this study.

\section{Related work}\label{related}

The measurement process has been addressed at length in SE. Fenton and Bieman’s well-known book \cite{fenton2014software}, along with many other publications that we will not cite here, is proof of this. However, the impact of the measurement process on practice has not received very much attention. This contrasts with other disciplines, such as medicine, where studies on the impact of measurers on measurement results (e.g., \cite{Gibbons2002inter, Spring2001intra}) are readily available. The Bland-Altman plot~\cite{Bland1986} used in Section~\ref{sec:eval} was originally developed to compare measurement instruments in medicine. There are many others.

Focusing on SE, there were fairly early warnings about the impact of measurement on experimentation, such as \textit{"The results of an experiment can be no more valid than the measurement of the constructs investigated. […] Many studies have elaborately defined the independent variables (e.g., the software practice to be varied) and hastily employed a handy but poorly developed dependent measure (criterion). Results from such experiments, whether significant or not, are difficult to explain"} ~\cite{curtis1980measurement}. Kitchenham et al.~\cite{kitchenham1995towards} states that it is necessary to critically review and evaluate all the research in this field. 

In the following years, however, measurement problems received less attention. One possible reason, according to Abran et al. \cite{Abran2003metrology}, is that measurement in SE has been carried out mainly from the viewpoint of mathematical measurement theory. This has meant that key measurement aspects, like measurement method and measurement instrument, have remained in the background. They defend the adoption, in SE \cite{VIM}, of practices proper to metrology, which do account for the above aspects.

Some SE authors have shown concern for the practical problems of measurement. Kitchenham et al.~\cite{kitchenham2001modeling} signal that, apart from relying on formally well-defined variables, the measurement process needs to be underpinned by a measurement protocol that \textit{"defines the who, when, and how of any data collection activity, i.e., (1) Who: the role responsible for data extraction; (2) When: the point in the development process when the measure should be taken; (3) How: what tools, methods are used to extract, record and store the data values"}. (We substitute the bullet items for numbers in the citation.) In a later, well-known article, Kitchenham et al.~\cite{Kitchenham2002} reiterate, perhaps less clearly, the above recommendations. 

Morasca~\cite{morasca2001software} suggests that, when the measurement process involves people, this process should, whenever possible, be automated. In the context of the measurement of response variables, Ko et al.~\cite{ko2015practical} indicate that there is a natural variation, which must be retained, as well as an unnatural, irregular, atypical variation (including everything that would not normally occur in the world and is due to the artificiality of the experiment), which must be eliminated.

García et al.~\cite{garcia2007managing} report a metamodel for defining software measurement models, which is a method for measuring any type of software entity using a set of generic metrics. However, their study is not concerned with aspects related to methods and instruments.

Gomez et al.~\cite{gomez2006systematic} report a systematic review of measurement in SE. In this study, they conclude that only 46\% of measurements have been empirically validated and again that only 28\% have been both empirically and theoretically validated. Khoshgoftaar et al.~\cite{khoshgoftaar1994applications} arrive at a similar conclusion within the more restricted information theory field.

Aversano and Tortorella \cite{aversano2018assessing} researched the impact of measurement tools on the evaluation of software maintainability metrics. The measurements show that the values of the metrics evaluated using measurement tools differ for the same software system.

\section{Conclusions}\label{sec:conclusiones}

During the experimental process, the operationalization of the response variables of the experiments is highly dependent on the researcher. Researchers operationalize each in their own way, and there are no protocols or recommendations about how to carry out the operationalization process. We addressed this problem by conducting a quantitative and qualitative study, obtaining the results reported in Sections 5 and 6, respectively. We now outline the conclusions and future research work.

From the operationalization process viewpoint, the FS experiment is no different from other TDD experiments available in the literature. As there was no experiment measurement process, the first step of our research was to define a protocol, based on the procedures used in other experiments run. This measurement procedure was framed within the SE measurement theory by Fenton and Bieman \cite{fenton2014software}.

We identified a set of relevant measurement method components, such as test suites, development environment, measurers, source code intervention type, metric , test framework, code compilers, etc. We selected and studied the components that we believed to have most influence on the measurement results: test suites \cite{dieste20}, measurers and source code intervention based on the experience acquired during the execution of the ESEIL experiment \cite{juristo2016experiences}.

The biggest problem that we detected concerns the test suites. In principle, we thought that the source of the problem was ad hoc test case creation, which a more formal, EP-based test generation method would solve. However, we were wrong. On the one hand, the test suite creation method is not intrinsically bad; the question is whether the test suite matches the expectations of the researchers and experimental subjects. If the researchers and experimental subjects do not expect code exceptions to be explicitly controlled, an EP-based test suite will return much poorer results than an AH test suite generated overlooking the exceptions. On the other hand, test suites tend not to respect a basic metric property known as representational condition: codes with more functionalities should return proportionally higher QLTY values. This is not generally the case with an EP-based test suite.

The main recommendation that has come out of this research is that test suites (and measurement instruments generally) cannot be designed exclusively taking into account theoretical considerations and should be verified in practice, whenever possible in the framework of a reproducibility analysis \cite{dieste20}.

We were surprised to find that the measurers had little impact and intervention types, a moderate influence on statistical analyses. In truth, we expected the exact opposite. This takes the pressure off the measurers, as it does not appear to be necessary to establish specific controls to assure the validity of the data gathered by different measurers who somehow manipulate the code delivered by the experimental subjects. In any case, we recommend the use of blinding with respect to the code delivered by the subjects, measurement results and treatments in order rule out bias in decision making. The use of intervention type protocols is beneficial and, therefore, also advisable, for increasing the similarity of the measurements.

\subsection{Operationalization in the scientific literature}

After reviewing the related scientific literature, we found that most experimental papers do not offer detailed information on how the response variables were operationalized. One especially important aspect is that is missing from most experimental reports is the measurement process. For example, TDD experiments do not indicate how the measurement test suites were generated. Likewise, these test suites are not usually available for independent verification. Predictably then, other details present in many TDD (and generally SE) experiments, such as intervention type and measurer, are, in many cases, not even mentioned. All this means that there are important factors that are not being taken into account and may be having an influence on the experimental results. 

This research leaves a lot of open questions, which we were not able to address on the grounds of time and resources. We list below only the most important future lines, some of which we hope to take up in the near future:

\begin{itemize}
	
	\item The recommendations that we have provided were inferred from the results reported here. However, as this research clearly shows, inferences are fallible. To increase the robustness of the findings of this research, this study needs to be repeated on other experiments using different quality response variables.
	
	\item It remains to determine how an experimenter can decide whether a measurement instrument (for example, AH test suite or EP test suite) is reliable. The steps of the measurement instrument reproducibility analysis should also be protocolized.
	
	\item We recommended blinding to reduce measurer bias when applying different intervention types. However, another path for improving the validity of the measurements would be to use different measurers, intervention types and even test suites, and average the results using either simple or weighted means.  
	
	\item We plan to analyse other experimental instruments, such as development environment, measurer learning and measurement protocols, which are not mentioned in this study and could also influence the results.
	
\end{itemize}

\begin{acks}
	This research was funded by the Spanish Ministry of Science, Innovation, and Universities research, Spain grants PGC2018- 097265-B-I00 and the FINESSE project, Spain (PID2021-122270OB- I00). This research was also supported by the Madrid Region R\&D program, Spain (project FORTE, P2018/TCS-4314) and the SATORI-UAM project (TED2021-129381B-C21).
\end{acks}

\bibliographystyle{ACM-Reference-Format}
\bibliography{sample-base}

\appendix

\end{document}